\documentclass [letterpaper,12pt]{article}

\usepackage{amsmath, amsthm}
\usepackage{amssymb}
\usepackage{graphicx}
\usepackage{verbatim}

\newtheorem{theo}{Theorem}
\newtheorem{theo_sec}{Theorem}[section]
\newtheorem{defi}{Definition}[section]

\newtheorem{lemm}{Lemma}[section]

\newcommand{\ud}{\,\mathrm{d}}

\begin{document}

\title{\centering\Large{An $n$-to-$1$ Bidder Reduction for Multi-item Auctions\\
and its Applications}}
\date{}

\author{Andrew Chi-Chih Yao\thanks{Tsinghua University and the Chinese University of Hong Kong. Email: \texttt{andrewcyao@tsinghua.edu.cn}. This work was supported in part by NSFC grant 61033001, and the Danish National Research Foundation and the National Science Foundation of China (under grant 61061130540). Part of this research was done while the author was visiting the Simons Institute of UC Berkeley during January-April, 2014.}}
\maketitle{}

\begin{abstract}
In this paper, we introduce a novel approach for reducing the $k$-item $n$-bidder auction with additive valuation to $k$-item $1$-bidder auctions. This approach, called the \emph{Best-Guess} reduction, can be applied to address several central questions in optimal revenue auction theory such as the power of randomization, and Bayesian versus dominant-strategy implementations. First, when the items have independent valuation distributions, we present a deterministic mechanism called {\it Deterministic Best-Guess} that yields at least a constant fraction of the optimal revenue by any randomized mechanism. Second, if all the $nk$ valuation random variables are independent, the optimal revenue achievable in {\it dominant strategy incentive compatibility} (DSIC) is shown to be at least a constant fraction of that achievable in {\it Bayesian incentive compatibility} (BIC). Third, when all the $nk$ values are identically distributed according to a common one-dimensional distribution $F$, the optimal revenue is shown to be expressible in the closed form $\Theta(k(r+\int_0^{mr} (1-F(x)^n) \ud x))$ where  $r= sup_{x\geq 0} \, x(1 - F(x)^n)$ and $m=\lceil k/n\rceil$; this revenue is achievable by a simple mechanism called \emph{2nd-Price Bundling}. All our results apply to arbitrary distributions, regular or irregular.
\end{abstract}

\newpage

\section{Introduction}

Consider the multiple items auction problem, in which a seller wants to sell $k$ items to $n$ bidders who have private values for these items, drawn from some possibly correlated probability distributions. We are interested in studying {\it incentive compatible} mechanisms under which the bidders are incentivized to report their values truthfully. One major question is how to design such mechanisms which can maximize the expected revenue for the seller.

The single-item case ($k=1$) was resolved by Myerson's classic work \cite{Myerson1981} when the bidders' values for the item are independently distributed. The general multiple-item case ($k>1$) is provably harder (e.g., \cite{Chen2014}\cite{DDT2014}), and has in recent years been intensively studied in the literature.  In particular, when the inputs are discrete, much progress has been made on the efficient computation of the optimal revenue (e.g., \cite{CDWSTOC2012}\cite{CDWFOCS2012}\cite{CDWSODA2013}\cite{CHSODA2013}). Another direction is to design simple mechanisms for approximating optimal revenues in various settings (e.g., \cite{CHK2007}\cite{CHMS2007}\cite{H2009}\cite{Ronen2001}).
However, there remain important aspects of the multiple-item auction that are not well understood. Most of the known results put restrictions on the distributions (e.g., \cite{ BMP1963}\cite{CDWSODA2013}\cite{CHSODA2013}).  Also, the computational methods proposed typically find the optimal revenue by solving some mathematical programming problems, which do not yield mathematical formulas for the optimal revenue (or its approximation), even for relatively simple input distributions. To name some intriguing open questions: Is it possible to express the optimal revenue in terms of the valuation distributions elegantly, can the optimal revenue be achieved by some simple mechanism, and is the requirement of \emph{dominant strategy incentive compatibility} (DIC) much more stringent than \emph{Bayesian incentive compatibility} (BIC)?

In this paper, we introduce a novel approach for reducing $k$-item $n$-bidder auctions with additive valuations to $k$-item $1$-bidder auctions. This approach, called the \emph{Best-Guess} reduction, can be applied to address some of the above central questions in optimal revenue auction theory regarding the power of randomization, and Bayesian versus dominant-strategy. First, when the items have independent valuation distributions, we present a deterministic mechanism called {\it Deterministic Best-Guess} that yields at least a constant fraction of the best randomized mechanism. Second, if all the $nk$ valuation random variables are independent, the optimal revenue achievable in {\it dominant strategy incentive compatibility} (DSIC) is at least a constant fraction of that achievable in {\it Bayesian incentive compatibility} (BIC). Third, when all the $nk$ values are identically distributed according to a common one-dimensional distribution $F$, the optimal revenue can be expressed in the closed form $\Theta(k(r+\int_0^{mr} (1-F(x)^n) \ud x))$ where $r= sup_{x\geq 0} \, x(1 - F(x)^n)$ and $m=\lceil k/n\rceil$; this revenue is achievable by a simple mechanism called \emph{2nd-Price Bundling}. All our results apply to arbitrary distributions, regular or irregular.

{\it Related Work:} The reduction of mechanism design from an $n$-bidder multi-item auction to $1$-bidder multi-item auction was considered in Alaei \cite{Al2011} with a different approach, which did not yield constant factor approximation in the DSIC model (except under restrictions such as `budget-balanced cross monotonicity'; see also \cite{AlFu2012}). Recently, Hart and Nisan \cite{HN2012} started a line of research (see \cite{Baba2014}\cite{GK2014}\cite{HN2013}\cite{LY2013}\cite{TW2013}) for studying simple mechanisms for $1$-bidder $k$-item auctions with provable performance bounds for {\it arbitrary} distributions; some of these results will be needed in our paper. The question of how much {\it randomization} helps in auction mechanism design has been studied in a variety of models (e.g.,  \cite{B2010}\cite{CMS2010}\cite{HR2012}). In some situations, such as in the $1$-bidder case (Babaioff et al. \cite{Baba2014}) and in the $n$-bidder unit-demand setting (Chawla et al. \cite{CMS2010}), it is known that randomized mechanisms can yield at most a constant factor over deterministic mechanisms. The question of how much more revenue BIC implementation can yield over DSIC has a large literature (e.g.,  Gershkov et al. \cite{GG2007}, Manelli and Vincent \cite{MV2007}). For the one-dimensional models (i.e. $k=1$), starting with Myerson's classical work, strict equivalence between BIC and DSIC has been established in various contexts. It is widely agreed that strict equivalence is false for the multi-dimensional settings, but how much revenue can BIC yield over DSIC is largely unknown.

\section{Preliminaries}

\subsection{Basic Concepts}

Let $\mathcal{F}$ be a multi-dimensional distribution on $[0, \infty)^{nk}$.  Consider the $k$-item $n$-buyer auction problem where the valuation $n\times k$ matrix $x=(x_i^j)$ is drawn from $\mathcal{F}$.  Buyer $i$ has $x_i\equiv(x_i^1, x_i^2, \cdots, x_i^k)$ as his valuations of the $k$ items. For convenience, let $x_{-i}$ denote the valuations of all buyers except buyer $i$; that is, $x_{-i}=(x_{i'}\,|\,1\leq i'\neq i\leq n)$.

A \emph{mechanism} $M$ specifies an \emph{allocation} $q(x)=(q_i^j(x)) \in [0, \infty)^{nk}$, where $q_i^j(x)$ denotes the probability that item $j$ is allocated to buyer $i$ when $x=(x_i^j)$ is reported to $M$ by the buyers. We require that $\sum_{i=1}^n q_i^j(x)\leq 1$ for all $j$, so that the total probability of allocating item $j$ is at most $1$. $M$ also specifies a \emph{payment} $s_i(x)\in (-\infty, \infty)$ for buyer $i$.  A mechanism is called \emph{dominant-strategy individually rational} \emph{(DSIR)} if for each  $i$ and $x$, $\sum_{j=1}^k x_i^j q_i^j (x) - s_i(x)\geq 0$, i.e., a buyer gets at least as much in (reported) value as he pays for.  A mechanism is called \emph{dominant-strategy incentive compatible} \emph{(DSIC)} if for every $i, x_i, x_{-i}, x'_i$, $\sum_{j=1}^k x_i^j q_i^j (x_i, x_{-i}) - s_i(x_i, x_{-i})\geq \sum_{j=1}^k x_i^j q_i^j (x'_i, x_{-i}) - s_i(x'_i, x_{-i})$.  That is, buyer $i$ does not gain any more utility by mis-reporting $x_i$ as $x'_i$, given that all other buyers maintain their reported valuations.

We also consider a weaker version of rationality and incentive compatibility that is widely adopted. A mechanism is called \emph{Bayesian individual rational (BIR)} if each buyer $i$ gets at least as much value as he pays for in the average sense, when all other buyers report truthfully. More precisely, BIR requires that for every $i$ and $x_i$, $E_{x_{-i}}(\sum_{j=1}^k x_i^j q_i^j(x_i, x_{-i}) -s_i (x_i, x_{-i}))\geq 0$.  Similarly, a mechanism is called \emph{Bayesian incentive compatible (BIC)} if for every $i, x_i, x'_i$, $E_{x_{-i}}(\sum_{j=1}^k x_i^j q_i^j(x_i, x_{-i}) -s_i (x_i, x_{-i}))\geq E_{x_{-i}}(\sum_{j=1}^k x_i^j q_i^j(x'_i, x_{-i}) -s_i (x'_i, x_{-i}))$.

Let $s(x)=\sum_{i=1}^n s_i(x)$ be the total payments received by the seller. For any mechanism $M$ on $\mathcal{F}$, let $s_M(\mathcal{F})= E_{x\sim \mathcal{F}}(s(x))$ be the (expected) revenue received by the seller from all buyers. The \emph{optimal} revenue is defined as $REV(\mathcal{F})=\sup _M s_M(\mathcal{F})$ when $M$ ranges over all the DSIR and DSIC mechanisms. Similarly, in the Bayesian model, the optimal revenue is defined as $REV_{Bayesian}(\mathcal{F})=\sup _M s_M(\mathcal{F})$ when $M$ ranges over all the BIR and BIC mechanisms. A mechanism is said to be \emph{deterministic} if all $q^j_i(z)\in \{0,1\}$. Let $DREV(\mathcal{F})$ denote the $\sup$ of revenue over all deterministic  DSIR-DSIC mechanisms for distribution $\mathcal{F}$.

 One well-known DSIR and DSIC mechanism is the \emph{Vickrey 2nd-price} mechanism \cite{V1961} applied to each item.  That is, for each item $j$, the seller awards the item to the highest bidder (with any specified tie-breaking rule) on this item but at the 2nd highest bid price.  Let $X=(X_i^j)$ be the random variable matrix distributed according to $\mathcal{F}$.  Let $X^{j\textrm{[2nd]}}$ be the 2nd largest of $X_i^j$, $1\leq i\leq n$, and $X^{\textrm{[2nd]}}=\sum_{j=1}^k  X^{j\textrm{[2nd]}}$.  Then the revenue of this Vickrey 2nd-price mechanism $M_{vr}$ is $E_{x\sim \mathcal{F}}(X^{\textrm{[2nd]}})$, and hence $REV(\mathcal{F})\geq E_{x\sim \mathcal{F}}(X^{\textrm{[2nd]}})$. Let $B(X_{-i})$ denote $(Y^1, Y^2, \cdots, Y^k)$, where $Y^j=\max\{X^j_{i'}| i'\not=i\}$.  That is,  $B(X_{-i})$ is the maximum bid among all buyers except buyer $i$. (\emph{Remark}: We sometimes write lower case $x$ for the random variable $X$ when there is no confusion, or write $X_{\mathcal{F}}$ to emphasize its relationship with $\mathcal{F}$.)

\subsection{$\beta$-Exclusive Mechanisms and the $\beta$-Bundling}

In this subsection we restrict ourselves to $1$-buyer $k$-item auctions.  In this case DSIR=BIR and DSIC=BIC, and we can simply call them IR, IC.   For $1$-bidder $k$-item auctions, we introduce a concept called \emph{$\beta$-exclusive mechanisms} which will be central to our reduction method. Let $\mathcal{L}$ be any distribution\footnote{Throughout this paper, we use $\mathcal{L}$ to denote the value distribution for 1-bidder auction, and $\mathcal{F}$ the value distribution for $n$-bidder auction.} over $[0, \infty)^k$, and $\beta = (\beta^1, \beta^2, \cdots, \beta^k)$ a vector from $[0, \infty)^{k}$.

\begin{defi}  Given  $\mathcal{L}$ and $\beta$, a mechanism M is called $\beta$-\emph{exclusive} if $q_M^j(z)=0$ whenever $z^j\leq \beta^j$; that is, an item $j$ with bid equal to or below the threshold $\beta^j$ will not be allocated to the bidder. Let $REV^X(\mathcal{L},\beta)$ be $\sup_M(E_{z\sim \mathcal{L}}(s_M(z)))$ over all $\beta$-{exclusive} IR-IC mechanisms M, and we refer to $REV^X(\mathcal{L},\beta)$ as the \textbf{optimal} \textbf{$\beta$-{exclusive} revenue} for $\mathcal{L}$.
\end{defi}
 As an example, the familiar concept of Myerson's \emph{reserve price} for each item may be viewed as a special case of $\beta$-exclusion. Also, any mechanism $M$ can be easily converted into a $0$-exclusive mechanism with the same revenue by setting $q^j(z)$ to $0$ whenever $z^j=0$.

 We introduce the following $\beta$-exclusive mechanism, called $\beta$-Bundling.  This mechanism will be useful in providing a deterministic implementation of our reduction in Theorem 3. (We remark that bundling is a widely studied mechanism with many interesting variants, see e.g. \cite{JVM2007} \cite{MV2006} \cite{P1983} \cite{TS2012}.)

First consider, for any $\beta \in [0, \infty)^{k}$ and $w\geq 0$, the mechanism $M_{\beta, w}$ with allocation $q$ and payment $s$, defined as follows:
\begin{align*} \text{If } \sum_{j, z^j>\beta^j }(z^j-\beta^j)
\begin{cases} \geq w, \ \text{then} &s(z)=w+ \sum_{j, z^j>\beta^j }\beta^j,  \\
                                         &\text{$q^j(z)=1$  if $z^j> \beta^j$ and $q^j(z)=0$ otherwise};\\
              < w,  \ \text{then}  &s(z)=0,\\
                                         &\text{$q^j(z)=0$ for all $j$}.
\end{cases}
\end{align*}
Use $M_{\beta, w}(\mathcal{L})$ to denote its revenue $E_{z\sim \mathcal{L}}(s(z))$. (The parameter $w$ may be regarded as an additional surcharge that the mechanism imposes on any bundle.)
\begin{defi} For any $\beta \in [0, \infty)^{k}$, let $R(\beta)=\{(\beta,\overline w) | \,\overline w\geq 0 \}\cup \{(\overline\beta,0) | \,\overline \beta\geq \beta \}.$  Given distribution $\mathcal{L}$, the \textbf{$\beta$-Bundling} for $\mathcal{L}$ is defined to be the mechanism $M_{\overline\beta, \overline w}$, where $(\overline\beta, \overline w)$ is chosen\footnote{In case the $\sup$ is not achieved at any finite point, we then simply pick a point $(\overline\beta, \overline w)$ with revenue arbitrarily close to the $\sup$.} to maximize $M_{\overline\beta,\overline w}(\mathcal{L})$ over all $(\overline\beta, \overline w)\in R(\beta)$.  That is, $(\overline\beta, \overline w)=arg\max_{(\overline\beta, \overline w)\in R(\beta)}M_{\overline\beta, \overline w}(\mathcal{L})$.  We use $\text{Bund}(\mathcal{L}, \beta)$ to denote the revenue of the {$\beta$-Bundling} mechanism for $\mathcal{L}$.
\end{defi}
\begin{lemm} The $\beta$-Bundling for $\mathcal{L}$ is a deterministic IR-IC mechanism.
\end{lemm}
\noindent \textbf{Proof. } Immediate from the definition. \qed

It will be shown in Section 6 (Theorem 6.1) that, when $\mathcal{L}=L^1\times L^2\times \cdots \times L^k$, this bundling mechanism yields a constant fraction of $REV^X(\mathcal{L},\beta)$, the best revenue achievable by any $\beta$-exclusive mechanism.

\section{Main Results}

    We start by considering the $1$-bidder $k$-item auction. Let $\mathcal{L}$ be any distribution over $[0, \infty)^k$, and $\beta$ a vector from $[0, \infty)^{k}$. To provide a good benchmark for $REV^X(\mathcal{L}, \beta)$, we define below an \emph{adjusted revenue} for any general IR-IC mechanism M (not necessarily $\beta$-exclusive), where the portion of M's revenue from allocating low-value items (relative to $\beta$) is effectively discounted.

    \begin{defi} Let $\mathcal{L}$ and $\beta$ be given.  For any IR-IC mechanism $M$ with allocation $q_M$ and payment $s_M$, define its {$\beta$-adjusted revenue} for $\mathcal{L}$ as $E_{x\sim\mathcal{L}}(s_M(x, \beta))$, where $s_M(x, \beta)= s_M(x)-\sum_{j, \,x^j \leq\beta^j} q_M^j(x)x^j$.  Let $REV^A(\mathcal{L}, \beta)$ be $sup_M E_{x\sim\mathcal{L}}(s_M(x, \beta))$ over all IR-IC mechanisms $M$, and we refer to $REV^A(\mathcal{L}, \beta)$ as the \textbf{optimal $\beta$-adjusted revenue} for $\mathcal{L}$ .
\end{defi}

Note that, if $M$ is a $\beta$-exclusive mechanism, then its $\beta$-{adjusted} revenue is equivalent to its normal revenue. Our first theorem compares the optimal $\beta$-exclusive revenue $REV^X(\mathcal{L},\beta)$ with the optimal $\beta$-{adjusted} revenue $REV^A(\mathcal{L},\beta)$;
this result will play a crucial role in our $n$-bidder to $1$-bidder reduction.
\begin{theo}\label{Theorem1} \emph{[$\beta$-Exclusion Theorem] }\   For any $\mathcal{L}=L^1\times L^2\times \cdots \times L^k$ and $\beta\in [0, \infty)^{k}$, we have  $REV^X(\mathcal{L},\beta) \geq \frac{1}{8}  REV^A(\mathcal{L}, \beta).$
\end{theo}

We now propose a reduction called \textit{Best-Guess} for the $n$-bidder $k$-item auction $\mathcal{F}$.  Under this reduction, only the top bidder for each item may get the item.  The seller performs $n$ separate 1-buyer auctions as follows:  In the auction for bidder $i$,  the seller uses a $\beta$-exclusive mechanism to enforce the top-bid constraint, where $\beta =B(x_{-i})$, i.e., $\beta^j=\max\{x_{i'}^j | \, i'\not=i\}$.

\fbox{\begin{minipage}{0.8\textwidth}
\noindent\textbf{\emph{Best-Guess Reduction}} for distribution $\mathcal{F}$
\vskip 6pt
\noindent  Given the $n\times k$ bid matrix $x=(x_i^j)$ distributed according to $\mathcal{F}$, the seller conducts with each bidder $i$ a $1$-bidder $k$-item auction with $x_i=(x_i^j|1\leq j \leq k)$ as the bid; the seller uses an IR-IC revenue-optimal $B(x_{-i})$-{exclusive mechanism} with respect to the distribution $x_i\sim X_i|x_{-i}$.
\end{minipage}}
\vskip 8pt

It is clear that the Best-Guess Reduction is indeed a valid mechanism (i.e., with each item getting total allocation $\leq 1$) that is \emph{DSIR} and \emph{DSIC}. We will use $BGR(\mathcal{F})$ to denote its expected total revenue. In actually implementing \emph{Best-Guess}, we may employ a $B(x_{-i})$-{exclusive mechanism} that is {$\alpha$-approximate} for each buyer $i$ (rather than a truly optimal mechanism), that is, one that yields at least $1/\alpha$ of the the optimal $B(x_{-i})$-exclusive mechanism's revenue.
We refer to this modified version of Best-Guess as \emph{$\alpha$-approximate Best-Guess}.  The resulting mechanism (by choosing any $\alpha$-approximate mechanism for each buyer $i$) is clearly DSIR-DSIC; we call it a \textbf{$BGR_\alpha$-mechanism} and denote\footnote{Here we are slightly abusing the notation for the sake of brevity.  When the notation $BGR_\alpha(\mathcal{F})$ is used, the specific $\alpha$-approximate mechanism used for each buyer will be clear from context.} its revenue by $BGR_\alpha(\mathcal{F})$.  In particular, when $\alpha=1$, we have $BGR_1(\mathcal{F})=BGR(\mathcal{F})$, the revenue of the Best-Guess Reduction itself.

 Note that $BGR(\mathcal{F})$ may not always yield good revenue; for example, $BGR(\mathcal{F})=0$ when all valuations $x_i^j$ are equal to a constant $c$. However, by simply taking the better of $BGR(\mathcal{F})$ and $E_{x\sim \mathcal{F}}(X^{\textrm{[2nd]}})$ (which is the revenue of the 2nd-price Vickrey mechanism), one can show that a constant fraction of $REV(\mathcal{F})$ is guaranteed when the items are independent.


\begin{theo}\label{Theorem2}\ The Best-Guess Reduction is a \emph{DSIR-DSIC} mechanism. Furthermore, for any $\mathcal{F}=\mathcal{F}^1 \times \cdots \times \mathcal{F}^k$, we have $BG(\mathcal{F})\equiv\max\{BGR(\mathcal{F}), E_{x\sim \mathcal{F}}(X^{\textrm{[2nd]}}) \}\geq  \frac{1}{9}REV(\mathcal{F})$.
\end{theo}

Theorem 2 says that, when $REV(\mathcal{F})$ is much larger (say 9 times more) than what the Vickrey 2nd-price can produce, Best-Guess Reduction can extract the revenue more effectively. Also note that $BG(\mathcal{F})$ is realizable by the mechanism formally defined as follows:

\noindent \textbf{Mechanism BG:}
\begin{align*} \text{If $BGR(\mathcal{F})$}
 \begin{cases}\geq E_{x\sim \mathcal{F}}(X^{\textrm{[2nd]}}) \text{\ then use the Best-Guess Reduction;}\\
                <  E_{x\sim \mathcal{F}}(X^{\textrm{[2nd]}}) \text{\ then use Vickrey 2nd-price mechanism}.
 \end{cases}
\end{align*}
The following Corollary generalizes Theorem 2 to the $\alpha$-approximate version of Best-Guess.

\noindent\textbf{Corollary.} The revenue $BGR_\alpha(\mathcal{F})$ of any $\alpha$-approximate mechanism satisfies\\ $\max\{BGR_\alpha(\mathcal{F}), E(X_{\mathcal{F}}^{\text{[2nd]}}) \}\geq \frac{1}{8\alpha +1} REV(\mathcal{F})$.
\vskip 6 pt

Theorem 2 corresponds to the $\alpha=1$ case of the Corollary.  The revenue $\max\{BGR_\alpha(\mathcal{F}), E(X_{\mathcal{F}}^{\text{[2nd]}})\}$ can be realized, as before, by a \textbf{Mechanism $BGR_\alpha$} which identifies with the better (for distribution $\mathcal{F}$) between a given $BGR_\alpha$-mechanism  and the Vickrey 2nd-price mechanism.

\fbox{\begin{minipage}{0.8\textwidth}
\noindent\textbf{\emph{Deterministic Best-Guess Reduction (DBGR)}} for distribution $\mathcal{F}$
\vskip 6pt
\noindent Given the $n\times k$ bid matrix $x=(x_i^j)$, the seller conducts with each bidder $i$ a $1$-bidder $k$-item auction with $x_i=(x_i^j|1\leq j \leq k)$ as the bid; the seller uses the $\beta$-Bundling mechanism for distribution $\mathcal{L}$, where $\beta=B(x_{-i})$ and $\mathcal{L}$ is defined by $X_i|x_{-i}$.
\end{minipage}}
\vskip 10pt

\begin{theo}\ $DBGR$ is a deterministic \emph{DSIR-DSIC} mechanism. Furthermore, if $\mathcal{F}=\mathcal{F}^1 \times \cdots
\times \mathcal{F}^k$, then $DBGR$ is a $BGR_\alpha$-mechanism for $\mathcal{F}$ where $\alpha=8.5$.
\end{theo}

From Theorem 3 and Corollary to Theorem 2, we have $\max\{DBGR(\mathcal{F}), E(X_{\mathcal{F}}^{\text{[2nd]}}) \}\geq \frac{1}{69} REV(\mathcal{F})$. In other words, the following deterministic mechanism can realize revenue $\frac{1}{69} REV(\mathcal{F})$:

\vskip 5pt
\noindent \textbf{Mechanism DBG:}
\begin{align*} \text{If $DBGR(\mathcal{F})$}
 \begin{cases}\geq E_{x\sim \mathcal{F}}(X^{\textrm{[2nd]}}) \text{\ then use the Deterministic Best-Guess Reduction;}\\
                <  E_{x\sim \mathcal{F}}(X^{\textrm{[2nd]}}) \text{\ then use a deterministic Vickrey 2nd-price mechanism}.
 \end{cases}
\end{align*}

\noindent\textbf{ Corollary.} $DREV(\mathcal{F})\geq \frac{1}{69} REV(\mathcal{F})$ if $\mathcal{F}=\mathcal{F}^1 \times \cdots \times \mathcal{F}^k$.

Turning next to a different question, we show that, if all the $nk$ value distributions are independent
(though not necessarily identical), then the optimal revenues are equivalent,
up to constant factor, for the Bayesian and the dominant strategy settings.
\begin{theo}\ If $\mathcal{F}=\otimes_{i,j} F_i^j$, then $REV_{\text{Bayesian}}(\mathcal{F})\leq 9\,REV(\mathcal{F})$.
\end{theo}
If all the $nk$ value distributions are independent and identical, we can obtain a formula in closed-form. We propose a new mechanism, called \emph{Second-Price Bundling (SPB)}, as a heuristic for approximating DBGR. Let a parameter $w \geq 0$ be first chosen.

\vskip 6pt
\fbox{\begin{minipage}{0.8\textwidth}
\noindent{\textbf{\emph{Second-Price Bundling (SPB)}}} with parameter $w$
\vskip 6pt
\noindent The SPB mechanism picks for each item a maximum bidder (breaking ties using the uniform random rule); let $J_i$ be the set of items for which bidder $i$ is the selected maximum bidder. For each $i$, the seller makes a take-or-leave offer for all the items in $J_i$ (as one bundle) to bidder $i$, at the price $w+\sum_{j\in J_i} x^{j\textrm{[2nd]}}.$
\end{minipage}}
\vskip 10pt

In this scheme, the parameter $w$ serves as a {surcharge} on top of the second price to enhance the revenue.  SPB can be regarded as a simplified version of DBGR, in which the Bundling mechanisms applied to different bidders have a common, fixed surcharge $w$.
  For any $\mathcal{F}$, let $SPB(\mathcal{F})$ be the maximum revenue that can be generated by any mechanism in the  SPB family (that is, over all possible choices of parameter $w$). In the $1$-bidder case, bundling is known
\cite{LY2013} to yield at least a constant fraction of the optimal revenue for iid items.  Theorem 5 shows that, for $n$ bidders, $SPB$ similarly achieves a constant fraction of the optimal revenue when all the $nk$ valuation random variables are iid according to a common one-dimensional distribution $F$. We denote such a valuation distribution $\mathcal{F}$ by $F^{n\otimes k}$.
\begin{theo} Let $\mathcal{F}=F^{n\otimes k}$, $r=\sup_{x\geq 0}x(1-F(x))$ and $m=\lceil k/n\rceil$.\\
(a) $REV(\mathcal{F}) = \Theta(k(r+\int_0^{mr} (1-F(x)^n) \ud x))$;\\
 (b) $SPB$ is an IR-IC mechanism for any chosen parameter value $w$, and $SPB (\mathcal{F}) =\Theta (REV(\mathcal{F}))$.
\end{theo}
We remark that the constants in the $\Theta$ notations in Theorems 5 are universal constants, i.e., independent of $n, k$, and ${F}$.
We will prove Theorems 1-5 in Sections 4-9. In addition to the above main results, Theorems 5.3, 6.1, 7.1 and 9.1 may also be of some independent interest. The Appendix contains the proofs of some auxiliary lemmas left out of the main text.

\section{Theorem 1: Effect of $\beta$-Exclusion}

In this section we prove Theorem 1. First some notations. Let $\mathcal{L}=L^1 \times \cdots \times L^k$ be a distribution over $[0, \infty)^k$, and $\beta = (\beta^1, \beta^2, \cdots, \beta^k) \in [0, \infty)^{k}$. Let $Y^j$ be the random variables corresponding to $L^j$. Define $\xi^j=Pr\{Y^j>\beta^j\}$.  For any real number $\alpha$, let $Y^j_\alpha$ be the random variable obtained as follows:  with probability $\xi^j$, generate $Y^j|(Y^j>\beta^j)$; otherwise let $Y^j_\alpha=\alpha$.
\vskip 6pt

\begin{defi} For any $u=(u^1, \cdots, u^k)$, let $\mathcal{Y}_u=Y^1_{u^1} \times\cdots \times Y^k_{u^k}$. Define $\mathcal{L}^+_\beta =\mathcal{Y}_\beta$, and  $\mathcal{L}^-_\beta =\mathcal{Y}_\gamma$  where $\gamma =(0, \cdots, 0)$.
\end{defi}
\begin{lemm} For any $u\in [0, \beta^1]\times \cdots \times [0, \beta^k]$, $REV^A(\mathcal{Y}_u,\beta) \leq REV^X(\mathcal{Y}_u, u).$
\end{lemm}
\noindent\textbf{Proof.} Let M be any IR-IC mechanism with allocation $q$ and payment $s$.  We construct mechanism $M'$ with allocation $q'$ and payment $s'$ defined by: for any $z$ in the support of $\mathcal{Y}_u$ and any $j$, let
\begin{align*} q'^j(z)&= \begin{cases} q^j(z) \text{\ \ if \ \ $z^j>u^j$}\\
                                       0 \text{\ \ \ \ \ \ otherwise,}
                         \end{cases} \\
               s'(z)&=s(z) - \sum _{j,\, z^j= u^j} u^j\cdot q^j(z).
\end{align*}
It is easy to verify that $M'$ is $u$-exclusive, IR-IC, and $E_{z\sim \mathcal{Y}_u}(s'(z)) = E_{z\sim \mathcal{Y}_u}(s(z, \beta))$.  This proves $REV^A(\mathcal{Y}_u,\beta) \leq REV^X(\mathcal{Y}_u, u)$, and hence Lemma 4.1.   \qed

\begin{lemm}$REV^A(\mathcal{L},\beta) \leq REV(\mathcal{L}_\beta^-).$
\end{lemm}
\noindent\textbf{Proof.} Clearly, with $\gamma=(0, \cdots, 0)$ we have
\begin{align*} REV(\mathcal{L}^-_\beta)= REV^A(\mathcal{Y}_\gamma,\beta).
\end{align*}
We will prove Lemma 4.2 in two steps:

\noindent \textbf{Step 1.} Prove that there exists $u\in [0, \beta^1]  \times \cdots \times [0, \beta^k]$ such that $REV^A(\mathcal{L}, \beta)\leq REV^A(\mathcal{Y}_u, \beta)$;\\
\noindent \textbf{Step 2.} Prove that $REV^A(\mathcal{Y}_u,\beta) \leq REV^A(\mathcal{Y}_\gamma, \beta)$.

For Step 1, we let $\mathcal{G}^{(0)}=\mathcal{L}$, and construct a sequence $u_1, u_2 \cdots$ inductively by choosing $u^j$ to maximize the value of $REV^A(\mathcal{G}^{(j+1)}, \beta)$ where $\mathcal{G}^{(j)}$ stands for $Y^1_{u^1} \times \cdots \times Y^{j-1}_{u^{j-1}}\times Y^{j}\times \cdots \times Y^{k}$.  We claim that, for each $j$,
\begin{align}REV^A(\mathcal{G}^{(j)}, \beta)\leq REV^A(\mathcal{G}^{(j+1)}, \beta).
\end{align}
If $\xi^j=1$ then for any choice of $u^j$, $Y^j_{u^j}=Y^j$, and hence Eq. 1 is true. We can thus assume
$\xi^j<1$.
Observe that $Y^{j}$ can be obtained as follows: Generate a random number $c$ distributed according to the distribution $Y^{j}|(Y^{j}\leq \beta^j)$ and then output a random number according to $Y^{j}_c$.  This immediately implies
\begin{align}REV^A(\mathcal{G}^{(j)}, \beta)= E_c(REV^A(Y^1_{u^1} \times \cdots \times Y^{j-1}_{u^{j-1}}\times Y^{j}_c\times Y^{j+1}\times\cdots \times Y^{k})).
\end{align}
Eq. 1 now follows from Eq. 2.
This finishes Step 1.

For Step 2, take an IR-IC mechanism M (with allocation $q$ and payment $s$) achieving $E_{z \sim \mathcal{Y}_u}(s(z, \beta))= REV^A(\mathcal{Y}_u, \beta)$. By Lemma 4.1, we can take M to be $u$-exclusive.

Consider a new mechanism $M'$ with allocation $q'$ and payment $s'$ satisfying $q'(z')= q(z)$ and $s'(z')= s(z)$, where $z$
is defined by $z^j=\max\{ z'^j, u^j\}$.  It is straightforward to verify that $M'$ is IR, IC and satisfies
$E_{z' \sim {\mathcal{Y}_\gamma}}(s'(z',\beta))=E_{z \sim {\mathcal{Y}_u}}(s(z,\beta))$.
This proves $REV^A(\mathcal{Y}_u, \beta)\leq REV^A(\mathcal{Y}_\gamma, \beta)$, and finishes Step 2. The proof of Lemma 4.2 is now complete. \qed
\begin{lemm}$REV^A(\mathcal{L}^+_\beta, \beta)\leq REV^X(\mathcal{L}, \beta).$
\end{lemm}
\noindent\textbf{Proof.} Take an IR-IC mechanism M  with allocation $q$ and payment $s$ achieving $E_{z \sim {\mathcal{L}^+_\beta}}(s(z, \beta))= REV^A(\mathcal{L}^+_\beta, \beta)$. By Lemma 4.1 and the fact $\mathcal{L}^+_\beta=\mathcal{Y}_\beta$, we can take M to be $\beta$-exclusive, and satisfying
\begin{align} E_{z \sim {\mathcal{L}^+_\beta}}(s(z))= REV^A(\mathcal{L}^+_\beta, \beta).
\end{align}
Now consider mechanism $M'$ with allocation $q'$ and payment $s'$ defined by: $q'(z')=q(z)$ and $s(z')=s(z)$ where $z^j=\max\{z'^j, \beta^j\}$.  It is easy to check that $M'$ is $\beta$-exclusive, IR-IC, and satisfies $E_{z \sim {\mathcal{L}}}(s'(z))= E_{z \sim {\mathcal{L}^+_\beta}}(s(z))$, and hence by Eq. 3 \[E_{z \sim {\mathcal{L}}}(s'(z))=REV^A(\mathcal{L}^+_\beta, \beta).\]  This proves Lemma 4.3.  \qed

It follows from Lemmas 4.2 and 4.3 that, to establish Theorem 1, it suffices to prove
\begin{align} REV(\mathcal{L}^-_\beta)\leq 8 REV^A(\mathcal{L}^+_{\beta},\beta),
\end{align}
to which we will devote the rest of this section.

Let  $M$, with allocation $q$ and payment $s$,  be an IR-IC mechanism achieving optimal revenue for distribution $\mathcal{L}_\beta^-$. It is well known (see \cite{HN2012}) that, without loss of generality we can assume M to have the NPT (no-positive-transfer) property, i.e., $s(z)\geq 0$ for all $z$. Furthermore, we can without loss of generality assume that M is $\gamma$-exclusive where $\gamma=(0, \cdots, 0)$, i.e., $q^j(z)=0$ whenever $z^j = 0$. (Otherwise, we can simply set $q^j(z)$ to $0$ whenever $z^j=0$.) We will construct a new mechanism $M'$ which, for distribution $\mathcal{L}_\beta^+$, has $\beta$-adjusted revenue on a par with $E_{z \sim {\mathcal{L}^-_\beta}}(s(z))$.

Let $D$ be the support of $\mathcal{L}^-_\beta$, that is, $D=({z^1}, {z^2},\cdots, {z^k})$ where $z^j \in (\beta^j, \infty)\cup \{0\}$.  The multi-set $\{(q(z), s(z))\,|\, z\in D\}$ can be considered as a \emph{menu} for M, so that the bidder with valuation $z$ can choose an entry $(q^*, s^*)$ from this set to maximize the utility $q^*z-s^*$.

To construct $M'$, we modify this menu by \emph{deleting} some entries and then \emph{lowering} the payment for all remaining entries.  Let $a>1$ and $0< b <1$ be two parameters satisfying $b > \frac{1}{a}$. A value $z\in D$ is said to be \emph{profitable} if
\begin{align} s(z)\geq a (\beta q(z))
\end{align}
where as usual $\beta q(z)= \sum_{j=1}^k \beta^j q^j(z)$.  Let $D_0\subseteq D$ be the set of all profitable values.  We construct for $M'$ the following menu:
\begin{align} \mathcal{M}'= \overline{\{(q(z), bs(z))|\, z\in D_0\}}
\end{align}
where we denote the \emph{closure} of a set $S\subseteq R^{k+1}$ by $\overline{S}$. Note that $\gamma\in D_0$, and hence $M'$ has an entry $(q(\gamma),  bs(\gamma))=(0,0)$ to ensure the IR property.

By definition of menu, the allocation $q'$ and payment $s'$ for $M'$ are determined as follows: For any bid $z' \in [0, \infty)^k$, let $q'(z')=u$ and $s'(z')= v$ where $(u,v)$ is chosen from entries in $\mathcal{M}'$ to maximize $uz'-v$.  Clearly, $M'$ is IR and IC.
We will show that $M'$ yields the desired $\beta$-adjusted revenue.

For each $z\in D$, let $\psi(z)=z'$ where $z'^j=\max\{z^j, \beta^j\}$. Intuitively, $E_{z \sim {\mathcal{L}^-_\beta}}(s(z))$ is approximated well by contributions from just the profitable values $z$. For such $z$, and its corresponding $z'=\psi(z)$ in the support of $\mathcal{L}^+_\beta$, we show that the effective payment $s'(z', \beta)$ is at least a fraction of the payment from the natural candidate entry $(q(z), b s(z))$ in $\mathcal{M}'$.  (We remark that the lowering of the payment in $\mathcal{M}'$ plays a crucial role in ensuring this property.) Let us define
\begin{align} c = 1-\frac{1}{ 1+a(1-b)}.
\end{align}
Clearly, $0< c< 1$.
\begin{lemm} Let $z\in D_0$ be a profitable value, then $s'(z', \beta)\geq (b-\frac{1}{a})c s(z)$ where $z'=\psi(z)$.
\end{lemm}

\noindent\textbf{Proof.} Suppose the lemma is false. Then there must exist in $\mathcal{M}'$ an entry $(q(u), b\, s(u))$ with $u\in D_0$ such that:
\begin{align} s'(z', \beta)< (b-\frac{1}{a})c s(z),
\end{align}
where $s'(z', \beta)=b s(u) - \sum_{j,\, z'^j=\beta^j} q^j(u)\beta^j$,
and
\begin{align}
q(z)z' - b s(z) \leq  q(u)z' - b s(u).
\end{align}
We derive a contradiction.

Note that as $u\in D_0$, we have $\sum_{j} q^j(u)\beta^j\leq\frac{1}{a}s(u).$  Thus
\begin{align} s'(z', \beta)\geq (b-\frac{1}{a})s(u).
\end{align}
It follows from Eqs. 8, 10 that
\begin{align} s(u)< c s(z).
\end{align}
Note that as M is IR and IC,
\begin{align} z(q(z) - q(u)) \geq s(z) - s(u).
\end{align}
From Eq. 9,
\begin{align*} 0 &\geq q(z)z'- q(u)z' - b s(z)+ b s(u) \\
                 &\geq q(z)z - q(u)z - \sum_{j} \beta^j q^j(u)- b s(z)+ b s(u).
\end{align*}
By Eq. 12 and the fact $u\in D_0$, we have then
\begin{align*} 0 &\geq s(z) - s(u)- \frac{1}{a}s(u)-b s(z)+ b s(u)\\
                 &= (1-b + \frac{1}{a})(c s(z)-s(u)),
\end{align*}
contradicting Eq. 11.  This proves Lemma 4.4.  \qed

Using Lemma 4.4, and the fact that $s'(z', \beta)\geq 0$ (see Eq. 10) for all $z'$, we obtain
\begin{align} E_{z' \sim {\mathcal{L}^+_\beta}}(s'(z', \beta))
                                        &=    E_{z \sim {\mathcal{L}^-_\beta}}(s'(\psi(z), \beta))\nonumber\\
                                        &\geq E_{z \sim {\mathcal{L}^-_\beta}}(\mathcal{I}_{z\in D_0} s'(\psi(z), \beta))     \nonumber\\
                                        &\geq E_{z \sim {\mathcal{L}^-_\beta}}(\mathcal{I}_{z\in D_0}(b-\frac{1}{a})c\, s(z))     \nonumber\\
                                        &=  (b-\frac{1}{a})c (E_{z \sim {\mathcal{L}^-_\beta}}(s(z))-
                                        E_{z \sim {\mathcal{L}^-_\beta}}(\mathcal{I}_{z\in D- D_0}s(z))).
\end{align}
Since $s(z)< a \sum_{j} \beta^j q^j(z)$ for $z\in D-D_0$, we have
\begin{align}E_{z \sim {\mathcal{L}^-_\beta}}(\mathcal{I}_{z\in  D- D_0}s(z))
                           &< a E_{z \sim {\mathcal{L}^-_\beta}}(\sum_{j} \beta^j q^j(z)) \nonumber\\
                           &=    a\sum_{j=1}^k \beta^j E_{z \sim {\mathcal{L}^-_\beta}}(\mathcal{I}_{z^j>0} q^j(z)) \nonumber\\
                           &\leq a\sum_{j=1}^k \beta^j Pr_{y \sim {L^j}}\{ y> \beta^j \}  \nonumber\\
                           &=    a\sum_{j=1}^k \beta^j \xi^j.
\end{align}
It follows from Eqs. 13 and 14 that
\begin{align*} E_{z \sim {\mathcal{L}^+_\beta}}(s'(z, \beta))
                  \geq (b-\frac{1}{a})c\, (E_{z \sim {\mathcal{L}^-_\beta}}(s(z))-a\sum_{j=1}^k \beta^j \xi^j).
 \end{align*}
This proves
\begin{align*} REV^A(\mathcal{L}^+_{\beta},\beta)
                  \geq (b-\frac{1}{a})c\, (REV(\mathcal{L}^-_\beta)-a \sum_{j=1}^k \beta^j \xi^j).
\end{align*}
 Hence
 \begin{align*}REV(\mathcal{L}^-_\beta)\leq \frac{1}{(b-\frac{1}{a})c} REV^A(\mathcal{L}^+_{\beta},\beta) + a \sum_{j=1}^k \beta^j \xi^j.
 \end{align*}
 Taking $a=4$, $b=\frac{3}{4}$ and $c=\frac{1}{2}$ , we obtain
  \begin{align*}REV(\mathcal{L}^-_\beta)\leq 4 REV^A(\mathcal{L}^+_{\beta},\beta) + 4 \sum_{j=1}^k \beta^j \xi^j.
  \end{align*}
  But the term $\sum_{j=1}^k \beta^j \xi^j$ is bounded by $REV^A(\mathcal{L}^+_{\beta},\beta)$, as pricing items at $\beta^j$ will yield $\beta$-adjusted revenue $\beta\xi$. This then completes the proof of Eq. 4, and hence the $\beta$-Exclusion Theorem. \qed

  \emph{Remarks:} In the above derivation, we have set the values of parameters $a, b, c$ to optimize the resulted bound.
\section{Proof of Theorem 2}

\subsection{An Upper Bound for Revenue}
It is of interest to compare $BG(\mathcal{F})$ with the revenue achieved by using a relaxed version of the Best-Guess Reduction (BGR).  Suppose in the description of BGR, one were to drop the requirement of $B(x_{-i})$-{exclusive} mechanisms, but use any general mechanism with optimal $B(x_{-i})$-{\emph{adjusted revenue}} (while everything else is kept the same).  The resulting revenue, denoted by $BG^A(\mathcal{F})$ is defined formally as follows.
\begin{defi}  Define $BG^A(\mathcal{F})$ to be $\sum_{i=1}^n E_{x_{-i}}(REV^A(X_i | x_{-i}\,, B(x_{-i}))).$
\end{defi}

Theorem 5.1 shows that the quantity $BG^A(\mathcal{F})$ provides a useful upper bound to $REV(\mathcal{F})$ for arbitrary distribution
$\mathcal{F}$. In the next subsection we will show that this bound is tight when the items are independent.
\begin{theo_sec}  Best-Guess Reduction and Mechanism BG are both {DSIR-DSIC} mechanisms. Furthermore, for any distribution $\mathcal{F}$, \[BG(\mathcal{F}) \leq REV(\mathcal{F})\leq BG^A(\mathcal{F})+ E(X_{\mathcal{F}}^{\text{[2nd]}}).\]
\end{theo_sec}
The rest of this subsection will be devoted to proving Theorem 5.1. It is obvious that both $BGR$ and  Mechanism BG are DSIR-DSIC mechanisms for solving the auction problem, and hence $BG(\mathcal{F}) \leq REV(\mathcal{F})$.
It remains to prove the upper bound $REV(\mathcal{F})\leq BG^A(\mathcal{F})+ E(X_{\mathcal{F}}^{\text{[2nd]}})$. Consider any mechanism M with allocation $q^j_i$ and payment $s_i$.  We will prove that its revenue satisfies
\begin{align}\sum_{i=1}^n E_{x\sim \mathcal{F}}(s_i(x))\leq \sum_{i=1}^n E_{x_{-i}}(REV^A(X_i|x_{-i}\, , B(x_{-i}))) + E(X_{\mathcal{F}}^{\text{[2nd]}}),
\end{align}
which is sufficient to establish the desired upper bound in Theorem 5.1.
For each $x$ and buyer $i$, let $\beta=B(x_{-i})$ and define
\begin{align}s_i(x, \beta)= s_i(x) -\sum_{j, x^j_i\leq \beta^j} q^j_i(x) x_i^j.
\end{align}
Then, noting that $x^j_i\leq \beta^j$ implies $x^j_i\leq x^{j \text{[2nd]}}$, we obtain
\begin{align*} \sum_{i=1}^n s_i(x) &\leq \sum_{i=1}^n s_i(x, B(x_{-i} )) + \sum_{i=1}^n\sum_{j=1}^k q^j_i(x) x^{j{\text{[2nd]}}} \\
                   &= \sum_{i=1}^n s_i(x, B(x_{-i})) + \sum_{j=1}^k x^{j{\text{[2nd]}}} \sum_{i=1}^n q^j_i(x) \\
                   &\leq \sum_{i=1}^n s_i(x, B(x_{-i})) + \sum_{j=1}^k x^{j{\text{[2nd]}}}                        \\
                   &= \sum_{i=1}^n s_i(x, B(x_{-i})) + x^{\text{[2nd]}}.
\end{align*}
This implies
\begin{align} \sum_{i=1}^n E_{x\sim \mathcal{F}}(s_i(x))\leq \sum_{i=1}^n E_{x\sim \mathcal{F}}(s_i(x, B(x_{-i})))+ E(X_{\mathcal{F}}^{\text{[2nd]}}).
\end{align}
Fix $i, x_{-i}$, and consider the induced IR-IC mechanism M' (for 1-bidder $k$-item auction) which, for bid $x_i\in [0,\infty)^k$, allocates $q^j_i(x_i, x_{-i})$ for item $j$ and gets payment $s_i(x_i, x_{-i})$. By Eq. 16 and the definition of $REV^A$, we have
\begin{align} E_{x_i}(s_i(x, B(x_{-i})))\leq REV^A(X_i|x_{-i}\,, B(x_{-i})).
\end{align}
Inequality 15 follows immediately from Eqs. 17 and 18.  This completes the proof of Theorem 5.1. \qed
\subsection{Optimality of Best-Guess}

When the items have independent valuation distributions, i.e. $\mathcal{F}=\mathcal{F}^1 \times \mathcal{F}^2 \times \cdots \times \mathcal{F}^k$, we show that the upper and lower bounds in Theorem 5.1 differ by at most a constant factor.

\begin{theo_sec}\label{Theorem5.2} Let $\mathcal{F}=\mathcal{F}^1 \times \cdots \times \mathcal{F}^k$, where $\mathcal{F}^j$ is item $j$'s valuation distribution over $[0, \infty)^n$.  Then
$BG^A(\mathcal{F})\leq 8\, BGR(\mathcal{F}).$
\end{theo_sec}
\noindent \textbf{Proof.} Observe that by definition of the Best-Guess Reduction, we have
\begin{align} BGR(\mathcal{F}) = \sum_{i=1}^n E_{x_{-i}}(REV^X(X_{i}|x_{-i}\,, B(x_{-i}))).
 \end{align}
 To prove Theorem 5.2, we need to express $BG^A(\mathcal{F})$ in similar form and compare it with Eq. 19.  By definition,
 \begin{align*} BG^A(\mathcal{F}) = \sum_{i=1}^n E_{x_{-i}}(REV^A(X_{i}|x_{-i} \,, B(x_{-i}))).
 \end{align*}
 Applying Theorem 1 with $\mathcal{L}=X_i|x_{-i}$, \,$\beta=B(x_{-i})$,  and using Eq. 19, we obtain
 \begin{align*} BG^A(\mathcal{F})
  &\leq 8 \sum_{i=1}^n E_{x_{-i}}(REV^X(X_{i}|x_{-i}\,,B(x_{-i})))\nonumber\\
                    &= 8 BGR(\mathcal{F}),
 \end{align*}
 and Theorem 5.2 is proved.  \qed

 By Theorems 5.1 and 5.2, we obtain
 \begin{align}REV(\mathcal{F})  &\leq  8 BGR(\mathcal{F}) + E(X_{\mathcal{F}}^{\text{[2nd]}})\nonumber\\
 &\leq 9 \max\{BGR(\mathcal{F}), E(X_{\mathcal{F}}^{\text{[2nd]}}) \}\nonumber\\
 &= 9 BG(\mathcal{F}).
 \end{align}
This completes the proof of Theorem 2.  \qed

The Corollary to Theorem 2 can be proved in a similar way. We modify Theorem 5.2 to read (and easily verifiable)
\[BG^A(\mathcal{F})\leq 8 \alpha \, BGR_\alpha (\mathcal{F}),\]
and Eq. 20 then becomes
\begin{align*}
REV(\mathcal{F})  &\leq 8\alpha BGR_\alpha (\mathcal{F}) + E(X_{\mathcal{F}}^{\text{[2nd]}})\nonumber\\
 &\leq (8\alpha +1) \max\{BGR_\alpha(\mathcal{F}), E(X_{\mathcal{F}}^{\text{[2nd]}}) \},
 \end{align*}
 proving the Corollary.
\vskip12pt

\begin{theo_sec}\label{Theorem5.3} Let $SREV(\mathcal{F})$ be the revenue obtained by selling each item separately and optimally.  If $\mathcal{F}={\mathcal{F}}^1 \times \cdots \times {\mathcal{F}}^k$, then \[SREV(\mathcal{F}) > \frac{c}{\log_2 (k+1)}REV(\mathcal{F})\] for some universal constant $c>0$.
\end{theo_sec}
\noindent Proof of Theorem 5.3 is done by using the Best-Guess Reduction and applying the $\frac{c}{\log_2 (k+1)}$-approximation result from \cite{LY2013} for the $1$-bidder $SREV$, and will be omitted here.

Theorem 5.3 strengthens results in \cite{Ronen2001} where such a bound was derived for the case $k=1$, in \cite{HN2012} for the case $k=2$, and in \cite{Baba2014} for the situation when all $nk$ valuation distributions are independent. Our result only requires that the items have
independent distributions.

\section{Deterministic Best-Guess Reduction}

 Before proving Theorem 3, we first establish some useful facts about $1$-buyer $k$-item auctions. For any distribution $L$ over $(-\infty, \infty)$ and $c \in (-\infty, \infty)$, let $L-c$ denote the distribution obtained from $L$ by shifting the origin from $0$ to $c$. That is,
$Pr_{z\sim L- c}\{z>y\} = Pr_{z\sim L}\{z> y+c \}$ for all $y$.

\begin{defi} For any distribution $\mathcal{L}={L}^1 \times \cdots \times {L}^k$ over $[0, \infty)^{k}$ and  $\beta\in [0, \infty)^{k}$, let $\mathcal{L}-\beta$ denote the distribution $(L^1-\beta^1) \times \cdots \times (L^k-\beta^k)$.
\end{defi}
 Recall that $\mathcal{L}^+_\beta$ is a distribution, with support
$[\beta^1, \infty)\times \cdots\times [\beta^k, \infty)$, derived from $\mathcal{L}$ as in Definition 4.1. Note that $\mathcal{L}^+_\beta -\beta$ is a distribution over $[0, \infty)^{k}$. The next lemma relates the optimal revenue achievable
by $\beta$-exclusive mechanisms to that achievable by general mechanisms (without the $\beta$-exclusive restriction).  Let $\xi(\mathcal{L})
=(\xi^1 (\mathcal{L}),  \cdots, \xi^k (\mathcal{L}))$ where $\xi^j (\mathcal{L})= Pr_{z^j\sim L^j}\{z^j>\beta^j\}$.
\begin{lemm} For any $\mathcal{L}={L}^1 \times \cdots \times {L}^k$ and $\beta$, $REV^X(\mathcal{L},\beta) \leq \xi(\mathcal{L})\beta + REV(\mathcal{L}^+_\beta -\beta).$
\end{lemm}
\noindent \textbf{Proof.}  Given in the Appendix.   \qed

We will use a recent result from Babaioff et al. \cite{Baba2014}.  Let $SREV(\mathcal{L})$ and $BREV(\mathcal{L})$ be the optimal revenue by selling separately
and Grand Bundling, respectively.
\begin{lemm} (\cite{Baba2014}) For any $\mathcal{L}={L}^1 \times \cdots \times {L}^k$,
\[REV(\mathcal{L}) \leq 7.5 \, \max\{SREV(\mathcal{L}), BREV(\mathcal{L})\}.\]
\end{lemm}
We now turn to the proof of Theorem 3.  It is obvious that $DBGR$ is a deterministic \emph{DSIR-DSIC} mechanism.  We will show that $DBGR$ is a $BGR_\alpha$-mechanism for $\alpha = 8.5$; that is, $\beta$-Bundling is
an $\alpha$-approximation to the ideal optimal $\beta$-exclusive mechanism.
\begin{theo_sec} For any $\mathcal{L}={L}^1 \times \cdots \times {L}^k$ and $\beta$,
\[Bund(\mathcal{L}, \beta)\geq \frac{1}{8.5}\, REV^X(\mathcal{L}, \beta).\]
\end{theo_sec}
\noindent \textbf{Proof.}  By Lemma 6.1,
\begin{align} REV^X(\mathcal{L}, \beta)\leq \sum_{j=1}^k \beta^j \xi^j (\mathcal{L})+ REV(\mathcal{L}^+_\beta -\beta).
\end{align}
It is easily seen that
\begin{align}\sum_{j=1}^k \beta^j \xi^j (\mathcal{L})=M_{\beta, 0}(\mathcal{L})\leq {Bund}(\mathcal{L}, \beta).
\end{align}
For the other term in Eq. 21, by Lemma 6.2, there exists $w\in [0, \infty)$ such that
\begin{align}REV(\mathcal{L}^+_\beta - \beta)\leq 7.5\,\max\{w\cdot Pr_{z\sim \mathcal{L}^+_\beta} \{\sum_{j=1}^k(z^j-\beta^j)\geq w\}, \, SREV(\mathcal{L}^+_\beta - \beta)\}.
\end{align}
Similar to Eq. 22, for any $\epsilon>0$, there exists $\overline{\beta}\geq \beta$ such that
\[ SREV(\mathcal{L}^+_\beta - \beta)-\epsilon \leq M_{\overline{\beta}, 0}(\mathcal{L})\leq {Bund}(\mathcal{L}, \beta).
\]
Thus, taking the limit $\epsilon \rightarrow 0$, we obtain
\begin{equation} SREV(\mathcal{L}^+_\beta - \beta)\leq {Bund}(\mathcal{L}, \beta).
\end{equation}
Also, it is clear that
\begin{align*} &Pr_{z\sim \mathcal{L}^+_\beta} \{\sum_{j=1}^k(z^j-\beta^j)\geq w\}\\
              =&Pr_{z\sim \mathcal{L}} \sum_{j, z^j>\beta^j}(z^j-\beta^j)\geq w\},
\end{align*}
which implies that
\begin{align} w\cdot Pr_{z\sim \mathcal{L}^+_\beta} (\sum_{j=1}^k (z^j-\beta^j)\geq w)
\leq Bund(\mathcal{L}, \beta).
\end{align}
With help of Eqs. 24 and 25, we obtain from Eq. 23
\begin{align} REV(\mathcal{L}^+_\beta - \beta)\leq 7.5\,{Bund}(\mathcal{L}, \beta).
\end{align}
It follows from Eqs. 21, 22 and 26 that
\begin{align*} REV^X(\mathcal{L}, \beta)\leq 8.5 \,Bund(\mathcal{L}, \beta).
\end{align*}
This proves Theorem 6.1.  \qed

Theorem 3 follows immediately from the Corollary to Theorem 2 (with $\alpha = 8.5$) and Theorem 6.1.

\section{A General Implementation of Best-Guess Reduction}

The DBG Mechanism studied in Section 6 is one special way of implementing the Best-Guess Reduction.  In this section we show that any $1$-buyer $k$-item mechanism $M$ with approximation ratio $\frac{1}{\alpha}$ for the revenue can be transformed into an $1$-buyer $k$-item \emph{$\beta$-exclusive} mechanism $M'$ yielding at least $\frac{1}{\alpha+1}$ of $REV^X(\mathcal{L}, \beta)$; then, combined with Corollary to Theorem 2, it gives an $n$-buyer $k$-item mechanism with approximation ratio $\frac{1}{8\alpha+9}$ .  Viewed in this light, the DBG Mechanism can be regarded as the special case where $M$ is the mechanism studied in \cite{Baba2014}.

\begin{defi} Given $\beta$ and $M$, a $1$-buyer $k$-item mechanism with allocation $q$ and payment $s$, we define a mechanism $\Phi_{\beta}(M)$ as follows.  First, convert $M$ into an $0$-exclusive IR-IC mechanism simply by setting $q^j(z)$ to $0$ whenever $z^j=0$; then

 \noindent (i) let $\Phi_{\beta, 1}(M)$ be the $1$-buyer $k$-item mechanism $M'$ with allocation $q'$ and payment $s'$ defined by: for any $z'\in [0, \infty)^{k}$, let $z\in [0, \infty)^{k}$ where $z^j=\max\{z'^j-\beta^j, 0\}$ for $1\leq j\leq k$, and let $q'(z')=q(z)$, $s'(z')=s(z)+ \beta q(z)$;

\noindent (ii) let $\Phi_{\beta, 2}(M)$ be the $1$-buyer $k$-item mechanism $M'$ with allocation $q'$ and payment $s'$ defined by: for any $z'\in  [0, \infty)^{k}$ and $1\leq j \leq k$, let $q'^j(z')=1$ if $z'^j>\beta^j$ and $0$ otherwise; let $s(z')=\sum_{j,\,z'^j>\beta^j} \beta^j$.

\noindent Define $\Phi_{\beta}(M)=\Phi_{\beta, 1}(M)$ if for input distribution $\mathcal{L}$, mechanism $\Phi_{\beta, 1}(M)$ yields more (expected) revenue than $\Phi_{\beta, 2}(M)$; otherwise let $\Phi_{\beta}(M)=\Phi_{\beta, 2}(M).$
\end{defi}

For $\alpha \geq 1$, we say that a mechanism M is an $\alpha$-approximate mechanism for $\mathcal{L}$
if  $M(\mathcal{L})\geq \frac{1}{\alpha} REV(\mathcal{L})$.
\begin{theo_sec} Let $\alpha \geq 1$ and $\mathcal{L}={L}^1 \times \cdots \times {L}^k$.  If M is an IR-IC $\alpha$\emph{-approximate} mechanism for $\mathcal{L}^+_\beta -\beta$, then the mechanism $M'=\Phi_{\beta}(M)$ is an IR-IC  $\beta$\emph{-exclusive} mechanism. Furthermore, $M'(\mathcal{L})\geq \frac{1}{\alpha+1} REV^X(\mathcal{L}, \beta)$.
\end{theo_sec}
\noindent \textbf{Proof. } It is straightforward to verify that, for $i=1$ and $2$, $\Phi_{\beta, i}(M)$ is IR, IC, and $\beta$-exclusive. It immediately follows that $\Phi_{\beta}(M)$ is also.  The proof of the last part of the theorem generalizes the proof of Theorem 6.1.  By Lemma 6.1,
\begin{align} REV^X(\mathcal{L},\beta) \leq \xi(\mathcal{L})\beta + REV(\mathcal{L}^+_\beta -\beta).
\end{align}
Let $M_1=\Phi_{\beta, 1}(M)$ and $M_2=\Phi_{\beta, 2}(M)$. Then by definition of $M_2$ and $M'$,
\begin{align} \xi(\mathcal{L})\beta= M_2(\mathcal{L}) \leq M'(\mathcal{L}).
\end{align}
To bound the other term in Eq. 27, note that as M is $\alpha${-approximate} for $\mathcal{L}^+_\beta -\beta$ we have
\begin{align} REV(\mathcal{L}^+_\beta -\beta) \leq \alpha\, M(\mathcal{L}^+_\beta -\beta).
\end{align}
Let M have allocation $q$ and payment $s$, and $M_1$ have allocation  $q'$ and payment $s'$.  Then by definition of $M_1=\Phi_{\beta, 1}(M)$ and $M'$, we have from Eq. 29
\begin{align} REV(\mathcal{L}^+_\beta -\beta)
&\leq \alpha\, E_{z\sim \mathcal{L}^+_\beta -\beta}(s(z))\nonumber\\
&\leq \alpha\, E_{z\sim \mathcal{L}^+_\beta -\beta}(s(z)+\beta q(z))) \nonumber\\
&=    \alpha\, E_{z\sim \mathcal{L}^+_\beta -\beta}(s'(z+\beta))  \nonumber\\
&=    \alpha\, E_{z'\sim \mathcal{L}^+_\beta}(s'(z'))  = \alpha\, E_{z'\sim \mathcal{L}}(s'(z')) = \alpha\, M_1(\mathcal{L}) \nonumber\\
&\leq  \alpha\, M'(\mathcal{L}).
\end{align}
It follows from Eqs. 27, 28 and 30 that
\begin{align*} REV^X(\mathcal{L}, \beta)\leq (\alpha+1) M'(\mathcal{L}).
\end{align*}
This proves Theorem 7.1.  \qed

\section{Bayesian vs. Dominant Strategy Revenue}

 We prove Theorem 4 in this section. We first establish an analogous result to Theorem 5.1 For the Bayesian setting of incentive compatibility, under the assumption of valuation independence among the bidders. Recall that $SREV(\mathcal{F})$ is the revenue obtained by selling each item separately and optimally.

\noindent \textbf{Theorem 8.1.}  Let $\mathcal{F}=\mathcal{F}_1 \times \cdots \times \mathcal{F}_n$, where each $\mathcal{F}_i$ is bidder $i$'s valuation distribution over $[0, \infty)^k$.  Then
\[REV_{\text{Bayesian}}(\mathcal{F}) \leq BG^A(\mathcal{F}) +SREV(\mathcal{F}).\]

\noindent \textbf{Proof.}
 Consider an optimal BIR-BIC mechanism $M$ for $\mathcal{F}$ with allocation $q_i^j(x)$, payment $s_i(x)$ from buyer $i$ and hence total payment $s(x)=\sum_{i=1}^n s_i(x)$. Our goal is to show that
\begin{align}E_{x\sim \mathcal{F}}(s(x)) \leq BG^A(\mathcal{F})+ SREV(\mathcal{F}).
\end{align}
For each $1\leq i \leq n$, consider the following $1$-buyer $k$-item mechanism $M_i$ for buyer $i$.  For any valuation $x_i\in [0, \infty)^k$, let the allocation of $M_i$ be $\bar{q}_i^j(x_i)= E_{x_{-i}}(q_i^j(x_i,x_{-i}))$, and its payment be $\bar{s}_i(x_i)=E_{x_{-i}}(s_i(x_i,x_{-i}))$, where $x_{-i}$ is distributed according to $\mathcal{F}_{-i}$, the product of $\mathcal{F}_{i'}$ for all $i'\neq i$. We first note an important property of $M_i$.
\vskip 5pt

\noindent\textbf{Property P1.} $M_i$ is IR and IC: for any $x_i, x'_i \in [0, \infty)^k$,
\begin{align*} x_i \bar{q}_i(x_i) - \bar{s}_i(x_i) &\geq 0 \nonumber\\
              x_i \bar{q}_i(x'_i) - \bar{s}_i(x'_i) &\leq x_i \bar{q}_i(x_i) - \bar{s}_i(x_i)
\end{align*}
where $x_i \bar{q}_i(x_i)$ stands for $\sum_{j=1}^k x_i^j\bar{q}_i^j(x_i)$.

This property follows directly from the fact that $M$ is BIR and BIC.

We are now ready to analyze the performance of $M$ in comparison with $BG^A(\mathcal{F})$, where by definition

\begin{equation}
BG^A(\mathcal{F})= \sum_{i=1}^n E_{x_{-i}} (REV^A(X_i, B(x_{-i})).
\end{equation}
(Recall $B(x_{-i})=( B^1(x_{-i}), \cdots, B^k(x_{-i}))$, where $B^j(x_{-i})$ is the value of the maximum of $\{ x^j_{i'} |i'\not=i\}$.)

Given any valuation $(x_i^j)$, we define for each $i$, $I_i(x)=\{j|\, x_i^j \leq B^j(x_{-i})\}$, and
\begin{align*}t_i(x_i,x_{-i})=\bar{s}_i(x_i) - \sum_{j \in I_i(x)}  \bar{q}_i^j (x_i)x^j_i.
\end{align*}
 Then
\begin{align}
E_{x\sim \mathcal{F}}({s}(x)) &=\sum_{i=1}^n E_{x\sim \mathcal{F}}(\bar{s_i}(x_i))\nonumber\\
                              &=\sum_{i=1}^n E_{x\sim \mathcal{F}}({t_i}(x_i, x_{-i}))+
                                E_{x\sim \mathcal{F}}(\sum_{i=1}^n \sum_{j\in I_i(x)} \bar{q}_i^j (x_i)x_i^j)
\end{align}
\begin{lemm} $\sum_{i=1}^n E_{x\sim \mathcal{F}}(t_i(x_i, x_{-i}))\leq BG^A(\mathcal{F}).$
\end{lemm}
\noindent \textbf{Proof.} By Property P1 and definition of $REV^A$, we have for each $i, x_{-i}$,
\begin{align*}E_{x_i\sim \mathcal{F}_i}(t_i(x_i, x_{-i}))\leq REV^A(X_i, B(x_{-i})).
\end{align*}
The lemma now follows from Eq. 32.  \qed
\begin{lemm} $E_{x\sim \mathcal{F}}(\sum_{i=1}^n \sum_{j\in I_i(x)} \overline q_i^j(x_i) x_i^j)\leq SREV (\mathcal{F}).$
\end{lemm}
\noindent \textbf{Proof.} Define $Q_i^j(x)=1$ if $j \in I_i(x)$ and $0$ otherwise.  Let
\begin{align*} \Psi(\mathcal{F})=E_{x\sim \mathcal{F}}(\sum_{i=1}^n \sum_{j\in I_i(x)} \overline q_i^j(x_i) x_i^j).
\end{align*}
Then
\begin{align*} \Psi(\mathcal{F})&=\sum_{i=1}^n \sum_{j=1}^k E_{x} (x_i^j Q_i^j(x) \overline q_i^j(x_i))\\
                                &=\sum_{j=1}^k \sum_{i=1}^n E_{x_{-i}} E_{x_i}(x_i^j Q_i^j(x) E_{z_{-i}} (q_i^j(x_i,z_{-i})))\\
                                &=\sum_{j=1}^k \sum_{i=1}^n E_{x_{-i}} E_{x_i}E_{z_{-i}}(Q_i^j(x_i,x_{-i} ) x_i^j q_i^j(x_i,z_{-i})).
\end{align*}
Relabel $x_i$ as $z_i$, and we have
\begin{align} \Psi(\mathcal{F})&=\sum_{j=1}^k \sum_{i=1}^n E_{x_{-i}} E_{z}(Q_i^j(z_i,x_{-i} ) z_i^j
                                                                                  q_i^j(z))\nonumber\\
                               &=\sum_{j=1}^k E_{z} (T_j(z)),
\end{align}
where
\[ T_j(z)=\sum_{i=1}^n E_{x_{-i}}(Q_i^j(z_i,x_{-i}) z_i^j q_i^j(z)).
\]
For convenience, we define $X^{j\textrm{[max]}}= \max\{X_i^j  |1\leq i\leq n \}$.

For any $j, z$, we have
\begin{align}T_j(z)
&= \sum_{i=1}^n q_i^j(z) z_{i}^j E_{x_{-i}} (Q_{i}^j(z_{i}, x_{-i}))\nonumber\\
&= \sum_{i=1}^n q_i^j(z) z_{i}^j  Pr_{x_{-i}} \{ \max \{ x_{i'}^j \, | \, i' \not= i\} \geq z_{i}^j \}\nonumber\\
&\leq \sum_{i=1}^n q_i^j(z) z_{i}^j  Pr_{x} \{ X^{j\textrm{[max]}} \geq z_{i}^j \}\nonumber\\
&\leq \sum_{i=1}^n q_i^j(z) REV(X^{j\textrm{[max]}})\nonumber\\
&\leq REV(X^{j\textrm{[max]}}),
\end{align}
where we have used the fact that $REV(X^{j{\text{[max]}}})=\sup_{y\geq 0} y(Pr\{X^{j{\text{[max]}}}\geq y\})$ is the optimal Myerson revenue for a single item with value distribution $X^{j{\text{[max]}}}=\max \{X_i^j |\, 1\leq i\leq n\}$.

Thus, from Eqs. 34 and 35,
\begin{align*}\Psi(\mathcal{F})\leq \sum_{j=1}^k REV(X^{j{\text{[max]}}}) \leq SREV(\mathcal{F}) .
\end{align*}
This proves Lemma 8.2. \qed

 From Eq. 33 and Lemmas 8.1, 8.2, we immediately obtain Eq. 31. This completes the proof of Theorem 8.1.  \qed

We can now prove Theorem 4.  Let  $\mathcal{F}=\otimes_{i,j} F_i^j$. By Theorem 5.2, we have
$ BG^A(\mathcal{F}) \leq 8 BGR(\mathcal{F})\leq 8 REV(\mathcal{F}).$ It then follows from Theorem 8.1 that
\begin{align*}
 REV_{\text{Bayesian}}(\mathcal{F})\leq BG^A(\mathcal{F})+ SREV(\mathcal{F}) \leq 9\, REV(\mathcal{F}),
 \end{align*}
which completes the proof of Theorem 4.

\section{Optimal Revenue in I.D.D. Case}

We will prove Theorem 5 in this section. We first give some notations and lemmas.
\begin{defi} For any distribution $F$ on $[0, \infty)$ and integer $\ell \geq 1$, define $r_F=\sup_{x\geq 0} x(1-F(x))$,
\[A_\ell (F)=r_F +\int_0^{\ell r_F} (1-F(x)) \ud x \text{\ \ \ \ and\ \ \ \ } C_\ell (F)=\int_0^{\ell r_F} x \ud F(x).\]
\end{defi}

\noindent\textbf{Property P2.} $r_F+ C_\ell (F)\leq A_\ell (F) \leq 2 r_F+ C_\ell (F)$ for any $\ell \geq 1$.
\vskip 5pt

The proof of Property P2 is given in the Appendix. The following lemma from Li and Yao \cite{LY2013} will also be useful.
\begin{lemm} \cite{LY2013} For any distribution $L$ over $[0,\infty)$ and integer $\ell$, $REV(L^{\otimes k})=\Theta(k\, A_k(L))$.
\end{lemm}

In the rest of this section, for any distribution $F$ on $[0,\infty)$, we reserve the symbols $\hat F$ to denote the distribution defined by $\hat F(x)= (F(x))^n$, and $m$ to denote $\lceil k/n \rceil$. We can restate Theorem 5(a) as follows:
\begin{align} \text{For\ \ \ } \mathcal{F}= F^{n \otimes k}, \ \  REV (\mathcal{F}) = \Theta (  k\,A_m(\hat F)).
\end{align}
In this section, for convenience we allow the value distributions to have support on $(-\infty, \infty)^k$ instead of $[0, \infty)^k$.
All the terms such as \emph{mechanisms}, \emph{IR, IC, REV,} etc. are defined exactly as previously.

Let $\mathcal{L}={L}^1 \times \cdots \times {L}^k$, where $L^j=L$ for a common distribution $L$ over $(-\infty, \infty)$.
Let $Y$ be a random variable distributed according to $L$.  Let $p=Pr\{Y>0\}$, and let $Z$ be the conditional distribution $Y|\, (Y>0)$.
\begin{lemm} \[REV(\mathcal{L})\leq \sum_{\ell=0}^k{k\choose \ell}p^\ell(1-p)^{k-\ell} REV(Z^{\otimes\ell}).\]
\end{lemm}
\noindent \textbf{Proof.} Obviously, we can assume $p>0$. For any $I\subseteq\{1, 2, \cdots, k\}$, $z\in [0, \infty)^k$, let $\overline I=\{1, \cdots, k\}-I$, $z^I=(y^i|i\in I)$ and $z^{\overline I}=(y^i|i\not\in I)$. Let M be any IR-IC mechanism for $\mathcal{L}$, with allocation $q$ and payment $s$.  We show that
\begin{align} E_{z\sim \mathcal{L}}(s(z))\leq \sum_{\ell=0}^k{k\choose \ell}p^\ell(1-p)^{k-\ell} REV(Z^{\otimes\ell}).
\end{align}
Fix any $I\subseteq\{1, 2, \cdots, k\}$, $z^{\overline I}\in (-\infty, 0]^{|\overline I|}$. We construct a mechanism $M'$ for valuation $z^{I}\in (0, \infty)^{|I|}$, with allocation $q'$ and payment $s'$ defined as follows. Let
\begin{align*}
 &q'^j(z^I)=q^j(z^I, z^{\overline I}) \text{\ \ for $j\in I$}, \\
              &s'(z^I)=s(z^I, z^{\overline I})-\sum_{j\in \overline I} z^j q^j(z^I, z^{\overline I}).
\end{align*}
Note that $s'(z^I) \geq s(z^I, z^{\overline I})$. It is straightforward to check that $M'$ is IR and IC.  By definition of $REV$, we have
\begin{align*}E_{z^I}(s'(z^I))\leq REV(Z^{\otimes |I|}).
\end{align*}
For a random $z\in \mathcal{L}$, let $I$ denote the random variable  corresponding to the set $\{j|\, z^j>0\}$. Then
\begin{align*}E_{z\sim \mathcal{L}}(s(z))&=E_I E_{z^{\overline I}}E_{z^I}(s(z))\\
&\leq E_I E_{z^{\overline I}}(E_{z^I}(s'(z)))  \\
&\leq E_I REV(Z^{\otimes |I|})\\
&=\sum_{\ell=1}^k {k\choose \ell}p^\ell(1-p)^{k-\ell} REV(Z^{\otimes\ell}).
\end{align*} This proves Eq. 37 and hence Lemma 9.2.  \qed

We will prove the upper and lower bounds of Theorem 5 in the next two subsections, respectively.
\subsection{Revenue Upper bound}

In this subsection we prove the upper bound in Theorem 5, i.e., showing that for distributions of the form $\mathcal{F}=F^{n\otimes k}$,
\begin{align} REV (\mathcal{F}) \leq O(k\,A_m(\hat F)).
\end{align}
To directly use the upper bound technique developed in Section 5 for the current purpose is possible, but it involves lengthy calculations.  Instead, we will use a variant of Theorem 5.1.  Let $\mathcal{F}$ be any distribution over $[0, \infty)^{nk}$ for the $n$-buyer $k$-item auction. First, as a counterpart of $BG^A(\mathcal{F})$, we define below $FX_\beta(\mathcal{F})$, which is a version of adjusted revenue but with a fixed adjustment $\beta$ for all bidders (rather than using $B(x_{-i})$-adjusted revenue for bidder $i$, as in Definition 5.1).

\begin{defi} For any fixed $\beta\in [0, \infty)^{k}$, let $FX_\beta(\mathcal{F})=\sum_{i=1}^n E_{x_{-i}}(REV((X_i-\beta) | x_{-i}))$.
\end{defi}

\begin{theo_sec} For any $n$-bidder $k$-item valuation distribution $\mathcal{F}$,  $REV(\mathcal{F})\leq FX_\beta(\mathcal{F})+ \|\beta\|$,  where $\|\beta\|=\sum_j\beta^j$.
\end{theo_sec}

\noindent \textbf{Proof.}  The proof follows from the same outline as the proof of Theorem 5.1 (but simpler), and will be omitted.  \qed

We next consider some simple properties of the distributions $\hat F(x)$ and $H_F(x)$ where $H_F(x)\equiv 1-F(x)$. Let $x_0\in [0, \infty)$ be the unique real number satisfying $H_F(x_0-)\geq \frac{1}{n}\geq H_F(x_0)$. Writing $H_{\hat F}(x)$ as $b(H_F(x))$, where $b(z)= 1-(1-z)^n$. Clearly, $b(z)$ is an increasing function of $z$, and one can easily verify that $b(1/n)\geq 1/e$.  Thus, $H_{\hat F}(x_0-)\geq b(1/n)\geq 1/e$. This shows
\begin{align}
 r_{\hat F}\geq x_0 H_{\hat F}(x_0-)\geq \frac{x_0}{e}.
\end{align}
 Without loss of generality, one can assume $H_F(x_0)\neq 0$. Otherwise, $REV (\mathcal{F})\leq k x_0$ and hence Eq. 38 is already satisfied due to Eq. 39.  Let $F_0$ denote the distribution $F$ conditioned on $x>x_0$:
\begin{align*} F_0(x)=
\begin{cases} 1-\frac{H_F(x)}{H_F(x_0)} & \text{for } x \geq x_0,\\
              0                  & \text{for } x < x_0.
\end{cases}
\end{align*}

Let $\beta=(\beta^1, \cdots, \beta^k)$, where $\beta^j=x_0$ for all $j$. By definition,
\begin{align} FX_\beta(\mathcal{F})=\sum_{i=1}^n REV(X_i-\beta).
\end{align}
Let $i\in \{1, \cdots, n\}$ be fixed, and let $p=H_F(x_0)\leq 1/n$.  Lemma 9.2 implies that
\begin{align*}
REV(X_i-\beta) &\leq \sum_{\ell = 0}^k {k\choose \ell} p^\ell(1- p)^{k-\ell} REV((F_0-x_0)^{\otimes\ell})\nonumber\\
&\leq \sum_{\ell = 0}^k {k\choose \ell} p^\ell(1- p)^{k-\ell} REV(F_0^{\otimes\ell})
\end{align*}
where we have used the elementary fact $REV((F_0-x_0)^{\otimes\ell}) \leq REV(F_0^{\otimes\ell})$.
By Lemma 9.1, we have $REV(F_0^{\otimes\ell})\leq c \,\ell A_{\ell}(F_0)$ for some constant $c>0$.  This leads to
\begin{align}
REV(X_i-\beta_i)\leq c \sum_{\ell = 0}^k {k\choose \ell} p^\ell(1- p)^{k-\ell} \ell A_{\ell}(F_0).
\end{align}

Eqs. 40 and 41 give an upper bound to $FX_\beta(\mathcal{F})$ in terms of $A_{\ell}(F_0)$. To derive Eq. 38, we only need to relate $A_{\ell}(F_0)$ to $A_{m}(\hat F)$.

\vskip 5pt
\noindent\textbf{Property P3.} Let $h=\lceil kp\rceil$ and $d_{n,k}(F)=\sum_{\ell=1}^k{k\choose \ell} p^\ell (1-p)^{k-\ell} \ell A_{\ell} (F_0)$. There exists a constant $c'>0$ such that for all $n, k, F$, we have $d_{n,k}(F)\leq c'{k}{p} A_h(F_0)$.
\vskip 5pt
\vskip 5pt
\noindent\textbf{Property P4.} For $h=\lceil kp\rceil$,  we have $A_h(F_0)\leq \frac{c''}{pn}  A_m(\hat F)$ where $c''=e + e \ln(2e)$.
\vskip 5pt

The proofs of Properties P3 and P4 are given in the Appendix. Using Properties P3 and P4, we obtain from Eqs. 40 and 41 that
\begin{align}
FX_\beta(\mathcal{F})&\leq c n
\sum_{\ell=
0}^k {k\choose \ell} p^\ell(1-p)^{k-\ell} \ell A_{\ell}(F_0)\nonumber \\
                       &= c n \,d_{n, k}(F) \nonumber\\
                       &\leq c c' nkpA_h (F_0) \nonumber\\
                       &\leq c c'c'' k A_m(\hat F).
\end{align}
By Theorem 9.1,  $REV (\mathcal{F})\leq FX_\beta(\mathcal{F})+ \|\beta\|$.
As $\|\beta\| = k x_0 \leq k\,e\,  r_{\hat F} \leq k\,e\, A_m(\hat F)$ by Eq. 39,
this together with Eq. 47 implies Eq. 42, completing the upper abound proof of Theorem 5.  \qed
\subsection{Optimality of Second-Price Bundling}

In this subsection, we prove that SPB is IR-IC and
\begin{equation} SPB(\mathcal{F}) \geq \Omega(k\,A_m(\hat F)) \text{\ \ where $m = \lceil k/n \rceil$}.
\end{equation}
Clearly SPB is IR, as it makes only take-or-leave offers.  We show that SPB is also IC.  Let $(x_i^j)$ be the true valuation.  If player $i$ reports his valuation truthfully, his utility $u_1$ is equal to $\sum_{j\in V}(x_i^j-z^j) -w$ where $z^j=\max \{x_{i'}^j |\,i'\neq i\}$ and $V=\{j |\, x_i^j> z^j\}$.  If player $i$ reports an untruthful valuation $y=(y^j |\, 1\leq j \leq k)$ for the $k$ items, then his utility $u_2$ will be $\sum_{j\in W}(x_i^j-z^j) -w$ for some $W\subseteq\{j | y^j\geq z^j\}$. Since $x_i^j-z^j \leq 0$ for all $j \not\in V$, we conclude that $u_2\leq u_1$, and player $i$ gains no advantage by reporting false valuation.  This proves SPB (with any parameter $w$) is IC.

 We next show that for some properly chosen $w$ (dependent on $\mathcal{F}$ only), SPB can achieve an expected revenue of  at least $\Omega(k A_{m}(\hat F))$.  We will choose the parameter $w$ as follows.   \\
 \textbf{Case 1.} $E(V_F)\geq \frac{1}{5} \max\{r_{\hat F}, \frac{C_m(\hat F)}{80}\}$.  In this case, by simply setting $w=0$, the SPB has exactly the same effect as selling each item separately with Vickrey's 2nd-price payment, yielding a revenue $k E(V_F)\geq \Omega(k A_m(\hat F))$. \\
 \textbf{Case 2.} $E(V_F) <  \frac{1}{5} \max\{r_{\hat F}, \frac{C_m(\hat F)}{80}\}$. In this case, pick some $u$ such that $u {H}_{\hat F}(u) \geq {4 \over 5} r_{\hat F}$. Define $q_0 =Pr\{W_F-V_F \geq \frac{u}{2}\}$, and define $w_0=u/2 $ if $mq_0\leq 1$ and $w_0=\lfloor q_0 m\rfloor u /2$ if $m q_0 >  1$. Choose the parameter $w$ to be
  \begin{align} w=
  \begin{cases} \frac{1}{4} m C_m(\hat F) \ \ &\text{if}\ \   C_m(\hat F)\geq 80\  r_{\hat F},\\
                w_0 \ \  &\text{if}\ \   C_m(\hat F)< 80\  r_{\hat F}.
   \end{cases}
    \end{align}

\begin{lemm} In Case 2, with the parameter $w$ as defined in Eq. 44, SPB can achieve expected revenue at least $\Omega(k A_{m}(\hat F))$.
\end{lemm}
The proof of Lemma 9.3 is given in the Appendix. Thus Eq. 43 is valid for all $\mathcal{F}$ (either Case 1 or Case 2). This completes the proof of Theorem 5.  \qed

\vskip 10pt
\noindent\textbf{\Large{Acknowledgements}} Thanks to Pingzhong Tang, Zihe Wang, and Song Zuo for their careful reading of the manuscript and pointing out some inaccuracies in an earlier version of the paper.

\noindent{\textbf{\Large{Appendix}}

\vskip 5pt
\noindent{\textbf{Note 1: Proof of Lemma 6.1}}

Let M be any IR-IC $\beta$-exclusive mechanism with allocation $q$ and payment $s$.  To establish the lemma, we prove
\begin{align*} E_{z\sim \mathcal{L}}(s(z)) \leq \xi(\mathcal{L})\beta + REV(\mathcal{L}^+_\beta -\beta), \tag{A1}
\end{align*}
where $\xi^j(\mathcal{L})=Pr_{z^j\sim L^j}(z^j>\beta^j)$, and $\xi(\mathcal{L})=(\xi^1(\mathcal{L}), \cdots, \xi^k(\mathcal{L})).$

\noindent{\textbf{Fact 1.}} $E_{z\sim \mathcal{L}}(s(z)) \leq \xi(\mathcal{L})\beta +E_{z\sim \mathcal{L}}(s(z)-q(z)\beta).$

\noindent{\textbf{Proof.}}  Simply note that $E_{z\sim \mathcal{L}}(s(z)) = E_{z\sim \mathcal{L}}(q(z)\beta)+ E_{z\sim \mathcal{L}}(s(z)-q(z)\beta)= \xi(\mathcal{L})\beta + E_{z\sim \mathcal{L}}(s(z)-q(z)\beta).$  \qed

We construct a mechanism $M'$ with allocation $q'$ and payment $s'$, so that it has a good expected revenue $s'(z')-q'(z')\beta$ for $z'\sim \mathcal{L}^+_\beta$.

For any $z\in [0, \infty)^k$, define $\psi(z)=z'$ where $z'^j=\max\{z^j, \beta^j\}.$  Clearly, $\psi(z)\in [\beta^1, \infty)\times\cdots\times [\beta^k, \infty).$ We now define $q', s'$.  For any $z'\in [\beta^1, \infty)\times\cdots\times [\beta^k, \infty)$, let $z=arg\max_z\{s(z)-q(z)\beta|\, z\in \psi^{-1}(z')\}$. Let $q'(z')=q(z)$ and $s'(z')=s(z)$.

\noindent{\textbf{Fact 2.}} $M'$ is IR-IC and
$ E_{z\sim \mathcal{L}}(s(z)-q(z)\beta)\leq E_{z'\sim \mathcal{L}^+_\beta}(s'(z')-q'(z')\beta)$.

\noindent{\textbf{Proof.}} For any $z'$ in the support of $\mathcal{L}^+_\beta$, i.e., $z'\in [\beta^1, \infty)\times \cdots \times [\beta^k, \infty)$, let $I(z')=\{j|\, z'^j > \beta^j\}$.  As M is $\beta$-exclusive, we must have for any $z\in \psi^{-1}(z')$, $q^j(z)=0$ if $j\not\in I(z')$.  This has several implications.  First, for any $z\in \psi^{-1}(z')$, $q(z)z'-s(z)= q(z)z-s(z)\geq 0$,
since M is IR.  This means $q'(z)z'-s'(z')\geq 0$, i.e., $M'$ is IR.

We now prove $M'$ is IC.  Define the function $u(y,z)=y q(z) - s(z)$, the utility under M obtained when the valuation is $y$ but reporting $z$.  We claim that for any $y, z \in \psi^{-1}(z')$, $u(y, z)=u(y, y)$.  To prove this claim, take M's IC condition $y(q(z)-q(y))\leq s(z)-s(y) \leq z(q(z)-q(y))$.  This leads to $s(z)-s(y)=\sum_{j\in I(z')}z'^j(q^j(z)-q^j(y))$, implying $u(z, z)=u(y, y)$. It is then straightforward to verify that $u(y, z)=u(y, y)$. In particular, take $y=z'$, and let $z=arg\max_z\{s(z)-q(z)\beta |\, z\in \psi^{-1}(z')\} $ as chosen in the construction of $M'$.  We have  $u(z', z)=u(z', z')$.  This means that the utility under $M'$ for valuation $z'$ remains the same as under M.  Thus, no advantage is gained by false reporting.  This proves that $M'$ is IC.

Finally, note that $z\in \psi^{-1}(z')$ implies $s(z)-q(z)\beta \leq s'(z')-q'(z')\beta$.  It follows that $ E_{z\sim \mathcal{L}}(s(z)-q(z)\beta)\leq E_{z'\sim \mathcal{L}^+_\beta}(s'(z')-q'(z')\beta)$.  This completes the proof of Fact 2. \qed

\noindent{\textbf{Fact 3.}} $ E_{z'\sim \mathcal{L}^+_\beta}(s'(z')-q'(z')\beta)\leq REV(\mathcal{L}^+_\beta -\beta).$

\noindent{\textbf{Proof.}} Define mechanism $M''$ for distribution $\mathcal{L}^+_\beta -\beta$ with allocation $q''$ and payment $s''$ by
\begin{align*}q''(z)=q'(z+\beta), \ \ \ \ s''(z)=s'(z+\beta)- \beta q'(z+\beta),
 \end{align*}
for all $z\in [0, \infty)^k$. It is straightforward to verify that  $M''$ is IR and IC.  It follows that
\begin{align*} E_{z\sim  \mathcal{L}^+_\beta}(s'(z)- \beta q'(z))&= E_{z\sim  \mathcal{L}^+_\beta  -\beta}(s'(z+\beta)- \beta q'(z+\beta))\\
   &= E_{z\sim  \mathcal{L}^+_\beta  -\beta}(s''(z))\\
   &\leq REV(\mathcal{L}^+_\beta -\beta).
 \end{align*}
 \qed

Facts 1-3 imply immediately Eq. A1.  This proves Lemma 6.1.

\noindent{\textbf{Note 2: Proof of Property P2}}

It is easy to see that
\begin{align*}C_\ell(F)&= - \int_0^{\ell r_F} x \ud (1- F(x)) = -x(1-F(x))|_0^{\ell r_F} + \int_0^{\ell r_F} (1- F(x)) \ud x\nonumber \\
&= - \ell r_F (1-F(\ell r_F)) + A_\ell(F)-r_F.
\end{align*}
Using the fact $1-F(\ell r_F)\leq \frac{r_F}{\ell r_F}=\frac{1}{\ell}$, we thus obtain  $0\leq A_\ell(F)-(r_F+C_\ell(F))\leq r_F$, and it follows that
\begin{align*} r_F+ C_\ell(F)\leq A_\ell(F) \leq 2 r_F+ C_\ell(F).
\end{align*}
This proves Property P2. \qed

\vskip 5pt
\noindent{\textbf{Note 3: Proof of Property P3}}

By definition, $A_{\ell}(F_0) = r_0 +\int_0^{\ell r}(1-F_0(x)) \ud x$, where $r_0=r_{F_0} = \sup_{x \geq 0} x(1-F_0(x)).$  For $\ell \geq h$,
\begin{align*} A_{\ell}(F_0) - A_{h}(F_0) &= \int_{h r}^{\ell r}(1-F_0(x)) \ud x\nonumber\\
&\leq \int_{h r}^{\ell r} \frac{r_0}{x} \ud x\nonumber = r_0 \ln\frac{\ell}{h}.  \tag{A2}
\end{align*}
Note that it implies, for all $\ell \leq 2 e h$,
\begin{align*}A_{\ell}(F_0)\leq A_{h}(F_0) + r_0 \ln (2 e).   \tag{A3}
\end{align*}
Using Eqs. A2, A3, we have
\begin{align*} d_{n,k}(F)\leq &\sum_{\ell = 1}^k {k\choose \ell} p^\ell(1- p)^{k-\ell}\cdot \ell (A_{h}(F_0) + r_0 \ln (2 e)) \\
&+r_0 \sum_{\ell > 2 e h } {k\choose \ell} p^\ell(1- p)^{k-\ell}\cdot \ell \ln \frac{\ell}{h}.
\end{align*}
Simplifying, we have
\begin{align*} d_{n,k}(F)\leq k p ( A_{h}(F_0) + r_0 \ln (2 e)) + r_0 \xi,  \tag{A4}
\end{align*}
where
\begin{align*} \xi= \sum_{\ell > 2 e h } {k\choose \ell} p^\ell(1- p)^{k-\ell} \cdot\ell \ln \frac{\ell}{h}.
\end{align*}
To estimate $\xi$, we use Stirling's approximation to obtain
\begin{align*} \xi &\leq \sum_{\ell > 2 e h } \frac{k^\ell}{\ell !}\cdot p^\ell\cdot \ell \ln \frac{\ell}{h}\\
                   &\leq \sum_{\ell > 2 e h } (\frac{kep}{\ell})^\ell\cdot \ell \ln \frac{\ell}{h}\\
                   &\leq e k p \sum_{\ell > 2 e h }\frac{\ln \ell}{2^{\ell -1}} \leq \beta_3 kp, \tag{A5}
\end{align*}
where $\beta_3$ is the constant $\sum_{\ell \geq 1} \frac{e \ln \ell}{2^{\ell -1}}$. Property P3 follows from Eqs. A4, A5. \qed
\vskip 5pt

\noindent\textbf{Note 4: Proof of Property P4}

We first derive a simple relation between $H_{\hat F}$ and $H$, where $H$ stands for $H_F$.

\noindent\textbf{Fact 4.} If $n H(x) \leq 1$, then $n H(x) \geq {H}_{\hat F}(x) \geq n H(x)/e.$

\noindent\textbf{Proof.} Assume $nH(x) \leq 1$, we show $nH(x) \geq H_{\hat F}(x)\geq \frac{1}{e} nH(x)$.  It is easy to check that $1-z\leq e^{-z} \leq 1 - z/e$ for all $z\in [0, 1]$.  It follows that $(1-H(x))^n \leq e^{-nH(x)} \leq 1- n H(x)/e$, implying $H_{\hat F}(x)=1-(1-H(x))^n\geq nH(x)/e.$  Furthermore, it is easy to verify that $(1-z)^n - 1 + nz$ is non-negative over $z\in [0, 1]$ by taking derivatives. This immediately leads to $n H(x) \geq H_{\hat F}(x)$ by letting $z= H(x)$.                        \qed

Let $H_0(x)$ stand for $H_{F_0}(x)=1-F_0(x)$, and $r_0$ stand for $r_{F_0}$. By definition, $H_0(x)=\frac{1}{p}H(x)$ for $x\geq x_0$, where $p=H(x_0)$.   It follows from $pn\leq 1$ and Fact 4 that, for $x\geq x_0$, $H_ {\hat F}(x)\leq H_0(x)\leq \frac{e}{pn} H_ {\hat F}(x)$; and these inequalities are easily checked to be true actually for all $x$.  This immediately implies
\begin{align*}
 r_{\hat F}\leq r_{0}\leq \frac{e}{pn} r_{\hat F}.   \tag{A6}
\end{align*}
If $kp\leq 1$, then $h=1$, and
\begin{align*}
 A_h(F_0)\leq 2 r_{0}\leq \frac{2e}{pn} r_{\hat F}\leq \frac{2e}{pn} A_m(\hat F),
\end{align*}
satisfying Property P5. We can thus assume $kp>1$.  In this case from A6
\begin{align*} hr_0\leq 2kp \frac{e}{pn} r_{\hat F}\leq 2 e m r_{\hat F}. \tag{A7}
\end{align*}
Hence, from A6-A7 we have
\begin{align*} \int^{h r_0}_0 H_0(x) \ud x &\leq \frac{e}{pn}\int^{2e m r_{\hat F}}_{0} H_{\hat F}(x) \ud x\\
  &=    \frac{e}{pn}(\int^{m r_ {\hat F}}_{0} H_ {\hat F}(x) \ud x+ \int^{2 e m r_ {\hat F}}_{m r_ {\hat F}} H_ {\hat F}(x) \ud x)  \\
  &\leq \frac{e}{pn}(\int^{m r_ {\hat F}}_{0} H_ {\hat F}(x) \ud x + \int^{2 e m r_ {\hat F}}_{m r_ {\hat F}} \frac{r_ {\hat F}}{x} \ud x)   \\
  &\leq \frac{e}{pn} (\ln(2e))A_m (\hat F).    \tag{A8}
\end{align*}
From Eqs. A6 and A8, we obtain
\begin{align*} A_{h}(F_{0})&= r_{0}+ \int^{h r_ {0}}_{0} H_0(x) \ud x\\
&\leq \frac{e}{pn} r_{\hat F} + \frac{e}{pn} (\ln(2e))A_m (\hat F) \\
&\leq \frac{c''}{pn} A_{m}(\hat F),
\end{align*}
This proves Property P4. \qed
\vskip 5pt

\noindent\textbf{Note 5: Proof of Lemma 9.3}

We first cite two earlier results from references [18] and [22].

\noindent\textbf{Fact 5.} [22] \ Let $m>1$ be any integer, and $G$ be a distribution on $[0, \infty)$ satisfying $C_m(G) > 10 r_G$.  Let $Z_1, \cdots, Z_m$ be $m$ iid distributions of $G$.  Then $ Pr\{\sum_{i=1}^m Z_i\geq \frac{1}{2} m C_m(G) \} \geq\frac{3}{4}.$
\vskip 5pt

\noindent\textbf{Fact 6.} [18] \ Let $Z_1, \cdots, Z_m$ be $m$ iid distributions of a one-dimensional distribution on $[0, \infty)$, and $t>0$ be any real number.  Define $w=t$ if $qm\leq 1$, and $w=t\lfloor qm\rfloor$ if $qm > 1$, where $q=Pr\{Z_i\geq t\}$.  Then $w\cdot Pr\{\sum_{i=1}^m Z_i\geq w\} \geq\frac{1}{4} m q t.$
\vskip 5pt

We also need the following estimate on a certain type of probability arising in our analysis.

\noindent\textbf{Fact 7.} Let $n, k > 1$ and $m = \lceil k/n\rceil$. Let $b_{n,k} = \sum_{\ell \geq m} {k\choose \ell} \frac{1}{n^\ell}(1- \frac{1}{n})^{k-\ell}$.  Then $b_{n,k}\geq \frac{k}{en}$ if $k\leq n$, and $b_{n,k}\geq \frac{1}{14}$ if $k> n$.
\vskip 5pt
\noindent\textbf{Proof.}
\noindent \textbf{Case 1:} $\frac{k}{n}\leq 1.$  Then $m=\lceil k/n\rceil = 1$, and $b_{n,k}\geq \frac{k}{n}(1- \frac{1}{n})^{k-1} \geq \frac{k}{n}(1- \frac{1}{n})^{n-1}$.  But $(1- \frac{1}{n})^{n-1}\geq \frac{1}{e}$ for all $n \geq 2$, hence $b_{n,k}\geq \frac{k}{en}$.

\vskip 5pt
\noindent \textbf{Case 2:} $\frac{k}{n}> 1.$  It is well known in statistics that the median lies to the left of the mean for binomial distributions.  Hence
\begin{align*}b'_{n,k}\geq 1/2   \tag{A9}
\end{align*}
where
\begin{align*}b'_{n,k}=\sum_{\ell\geq \lfloor k/n\rfloor}{k\choose\ell} \frac{1}{n^\ell}(1- \frac{1}{n})^{k-\ell}.   \tag{A10}
\end{align*}
We will show that
\begin{align*}b'_{n,k}\leq 7 b_{n,k},  \tag{A11}
\end{align*}
which together with Eq. A9 implies $b_{n,k}\geq \frac{1}{14}$, hence proving Fact 7.  We can assume $k/n$ to be non-integral; otherwise $b_{n,k}=b'_{n,k}.$  Note that from Eq. A10
\begin{align*}b'_{n,k}-b_{n,k} =  {k\choose{m-1}}\frac{1}{n^{m-1}}(1-\frac{1}{n})^{k-m+1} ,
\end{align*}
and
\begin{align*}b_{n,k} \geq {k\choose{m}}\frac{1}{n^m}(1-\frac{1}{n})^{k-m}.
\end{align*}
Thus, to prove Eq. A11 it suffices to show
\begin{align*}{k\choose{m-1}}\frac{1}{n^{m-1}}(1-\frac{1}{n})^{k-m+1}\leq 6 {k\choose{m}}\frac{1}{n^m}(1-\frac{1}{n})^{k-m}.
\end{align*}
This is equivalent to proving $n(1-1/n) \leq {6(k-m+1)/m}$, that is,
\begin{align*}(n-1)m \leq 6 (k-m+1), \ \ \ \textrm{or}\ \ \    (n+5)m \leq 6 (k+1).  \tag{A12}
\end{align*}
But Eq. A12 is easy to prove: using the inequality $n+{5k/n}\leq 5k +1$ for $n<k$, we have
\begin{align*}(n+5)m &=(n+5)\lceil k/n\rceil\leq (n+5)(\frac{k}{n} + 1)\\
&= n+ \frac{5k}{n}+k+5 \leq 6(k+1).
\end{align*}
This proves Eq. A11, hence the proof of Fact 7 is complete. \qed

We now proceed to prove Lemma 9.3. The proof is accomplished through a series of Facts. Recall that $m=\lceil \frac{k}{n}\rceil$, and it is assumed that
\begin{align*}
E(V_F)< \frac{1}{5} \max\{r_{\hat F}, \frac{C_m(\hat F)}{80}\}.  \tag{A13}
\end{align*}
Let ${J}=(J_1, J_2 \ldots, J_n)\in \mathcal{J}$. When $T_i=J_i$, player $i$ accepts the seller's offer if and only if $X^{J_i[\textrm{max}]}\geq X^{J_i[\textrm{2nd}]} + w$.  Let $P_w^{J_i}$ be the acceptance probability. Note that $P_w^{J_i}\leq P_w^{J'_i}$ if $J_i\subseteq J'_i$, since $J_i$ can be embedded as the first $|J_i|$ items of $J'_i$ without decreasing the acceptance probability.  Also, $P_w^{J_i}$ can be written as $P_w(|J_i|)$, as it depends only on $w$ and the cardinality of $J_i$.

\vskip 10pt
\noindent\textbf{Fact 8.} For any $1\leq i\leq n$, $E_{x\sim \mathcal{F}}(s_i(x)) \geq Pr\{|T_i|\geq m\} P_w(m)\cdot w.$

\noindent\textbf{Proof.} Immediate from the definitions and monotonicity of $P_w(m)$. \qed
\vskip 5pt
\noindent\textbf{Fact 9.} For any $1\leq i\leq n$,
\begin{align*} Pr\{|T_i|\geq m\} \geq
\begin{cases}
  \frac{k}{en} & \text{if } k\leq n,\\
  \frac{1}{14} & \text{if } k> n. \\
\end{cases}
\end{align*}

\noindent\textbf{Proof.} Note that $Pr\{|T_i|\geq m\}=\sum_{m'\geq m} {k\choose {m'}} \frac{1}{n^{m'}}(1-\frac{1}{n})^{k-m'}$. Fact 9 follows from Fact 7. \qed

Let $Y_1, Y_2 \ldots, Y_n$ be iid each distributed as $F$. Define $W_F=\max_i Y_i$ and $V_F=\textrm{2nd max of all}\  Y_i$.  Call $(W_F, V_F)$ a \textit{canonical pair}. Let $|J_i|=m$.  One can write $X^{J_i[\textrm{max}]}=\sum_{j=1}^m W_j$ and $X^{J_i[\textrm{2nd}]}=\sum_{j=1}^m V_j$, where $(W_1, V_1), \ldots, (W_m, V_m)$ are $m$ iid canonical pairs of random variables.  Then $P_w(m)$ can be written as the probability of the event
\begin{align*} \sum_{j=1}^m W_j\geq \sum_{j=1}^m V_j + w.  \tag{A14}
\end{align*}
\indent Recall that we have chosen the parameter $w$ as defined in Eq. 49 of the main text.

\noindent\textbf{Fact 10.} Let $D_m(\hat F)=\max \{ r_{\hat F}, \frac {1}{80}C_m(\hat F)\}$. Then  $w P_m(w)\geq \frac{1}{20}m D_m(\hat F)$.

\noindent\textbf{Proof.} We consider two separate cases.

\noindent\textbf{ Case A.}  \ \ $C_m(\hat F)\geq 80\ r_{\hat F}$.

We have $w=\frac{1}{4} m C_m(\hat F)$. As $C_m(\hat F)\leq mr_{\hat F}$ obviously, we have $m>1$. From Fact 5, we have
\begin{align*} a=Pr\{W_1+W_2+\cdots+W_m\geq 2w\}\geq \frac{3}{4}.
\end{align*}
Also, by Markov's Inequality and Eq. A13,
\begin{align*} b=Pr\{V_1+V_2+\cdots+V_m\geq w\}
&\leq Pr\{V_1+V_2+\cdots+V_m\geq 2 m E(V_F)\}\leq \frac{1}{2}.
\end{align*}
Thus,
\begin{align*} P_m(w)& =Pr\{\sum _{i=1}^m W_i \geq \sum _{i=1}^m V_i + w\}
\geq a-b \geq \frac{1}{4},
\end{align*}
and
\begin{align*} w P_m(w)  \geq \frac{1}{16} m C_m(\hat F)> \frac{m}{20} D_m(\hat F),
\end{align*}
as required.

\noindent\textbf{ Case B.}  \ \ $C_m(\hat F)< 80 r_{\hat F}$.

We first derive some useful information (Eq. A17 below).  In Case B, we have from Eq. A13
\begin{align*} E(V_F) < \frac{1}{5} r_{\hat F},  \tag{A15}
\end{align*}
and $u, q_0, w$ satisfying the conditions
\begin{align*} u H_{\hat F}(u) \geq \frac{4}{5} r_{\hat F},  \tag{A16}
\end{align*}
\begin{align*} q_0=Pr\{W_F-V_F \geq \frac{u}{2} \},
\end{align*}
\begin{align*} w=\begin{cases} {u}/{2}, \text{\ \ \ \ \ \ \ \ \ \ if $m q_0\leq 1$ ;}\\
                 \lfloor m q_0 \rfloor {u}/{2},  \text{\ \ if $m q_0> 1$.}
                 \end{cases}
\end{align*}
It follows from Eqs A15, A16 and Markov's Inequality that
\begin{align*} Pr\{V_F \leq \frac{u}{2} \}\leq \frac{E(V_F)} {u/2} < \frac{\frac{1}{5} r_{\hat F}} {\frac{2}{5}\frac{r_{\hat F}} {H_{\hat F}(u)}} = \frac{1}{2} H_{\hat F}(u).
\end{align*}
This implies
\begin{align*}   q_0\geq Pr\{W_F > {u} \} - Pr\{V_F \leq \frac{u}{2} \} \geq \frac{1}{2} H_{\hat F}(u).      \tag{A17}
\end{align*}
We are now ready to analyze $w P_m(w)$ for Case B. Apply Fact 6 with $Z_j=W_j-V_j$ and $t=\frac{u}{2}$, we have
\begin{align*} w P_m(w) = w\cdot Pr\{\sum _{j=1}^m (W_j- V_j)\geq w\}\geq\frac{uq_0 m}{8}.  \tag{A18}
\end{align*}
From Eqs. A16-A18,
\begin{align*} w P_m(w) \geq \frac{m u}{8}\cdot \frac{H_{\hat F}(u)}{2}\geq \frac{m}{16}\frac{4}{5}r_{\hat F}= \frac{m}{20}r_{\hat F}.
\end{align*}
 The assumption of Case B implies $r_{\hat F} = D_m(\hat F)$, thus we obtain $w P_m(w) \geq \frac{1}{20} m D_m(\hat F)$.  This completes the proof of Fact 10. \qed

Now, Facts 8-10 together imply that for $k\leq n$ (hence for $m=1$),
\begin{align*} E_{x\sim \mathcal{F}}(s_i(x)) \geq \frac{ k}{20 en}D_m(\hat F).
\end{align*}
Similarly, for $k>n$,
\begin{align*} E_{x\sim \mathcal{F}}(s_i(x)) \geq \frac{1}{280}m D_m(\hat F)
\geq  \frac{1}{280} \frac{k}{n} D_m(\hat F).
\end{align*}
This implies that, by choosing $w$ either way as in Eq. 44 (of the main text), we have
\begin{align*} E_{x\sim \mathcal{F}}(s(x))
 =\sum_{i=1}^n  E_{x\sim \mathcal{F}}(s_i(x)) \geq c k  D_m(\hat F)\geq \Omega(kA_m(\hat F))
\end{align*}
where $c=\frac{1}{280}$. Hence the proof of Lemma 9.3 is complete. \qed

\end{document}